\documentclass[prd,preprint,tightenlines,floatfix,
preprintnumbers,nofootinbib,eqsecnum,superscriptaddress]{revtex4}

\usepackage[T1]{fontenc}		

\usepackage{amsmath,amsfonts,amssymb,amstext,mathrsfs}
\usepackage{mathpazo}

\usepackage[dvips]{graphicx}
\usepackage{epsf,float}
\usepackage{revsymb}

\usepackage{dcolumn}
\usepackage{braket}
\usepackage{color,xcolor}
\usepackage{graphicx}
\usepackage{subfigure}
\usepackage{multirow}
\usepackage{tabularx}
\usepackage{pstricks}
\usepackage[section]{placeins}
\usepackage{booktabs}
\usepackage{array}

\usepackage{hyperref}

\newcommand{\Pom}{\mathbb{P}}

\newcommand{\Reg}{\mathbb{R}}

\newcommand{\bdPt}{\mbox{\boldmath $dP_{t}$}}
\newcommand{\bqta}{\mbox{\boldmath $q_{t,1}$}}
\newcommand{\bqtb}{\mbox{\boldmath $q_{t,2}$}}
\newcommand{\bpta}{\mbox{\boldmath $p_{t,1}$}}
\newcommand{\bptb}{\mbox{\boldmath $p_{t,2}$}}

\usepackage[normalem]{ulem}  

\begin{document}

\title{Study of the exclusive reaction 
$pp \to pp K^{*0} \bar{K}^{*0}$:\\
$f_{2}(1950)$ resonance versus diffractive continuum}
\vspace{0.6cm}

\author{Piotr Lebiedowicz}
\email{Piotr.Lebiedowicz@ifj.edu.pl}
\affiliation{Institute of Nuclear Physics Polish Academy of Sciences, Radzikowskiego 152, PL-31342 Krak{\'o}w, Poland}

\begin{abstract}
We present first predictions of the cross sections
and differential distributions for the exclusive reaction
$pp \to pp K^{*0} \bar{K}^{*0}$ contributing 
to the $K^{+} K^{-} \pi^{+} \pi^{-}$ channel.
The amplitudes for the reaction are formulated
within the nonperturbative tensor-pomeron approach.
We consider separately the $f_{2}(1950)$ $s$-channel exchange
mechanism and the $K^{*0}$ $t/u$-channel exchange mechanism,
focusing on their specificities.
First mechanism is a candidate for 
the central diffractive production of tensor glueball
and the second one is an irreducible continuum.
We adjust parameters of our model,
assuming the dominance of pomeron-pomeron fusion,
to the WA102 experimental data.
We find that including the continuum contribution 
alone one 
can describe the WA102 data reasonably well.
We present predictions for the reaction
$pp \to pp (K^{*0} \bar{K}^{*0} \to K^{+} K^{-} \pi^{+} \pi^{-})$ 
for the ALICE, ATLAS, CMS and LHCb experiments
including typical kinematical cuts.
We find from our model a cross sections of
$\sigma \cong 17-250$ nb for the LHC experiments,
depending on the assumed cuts.
Absorption effects are included in our analysis.
\end{abstract}

\maketitle

\section{Introduction}
\label{sec:intro}

Studies of the $K^{*} \bar{K}^{*}$ system 
have been carried out in two-photon interactions 
\cite{Albrecht:1987jp,Albrecht:1988gr,Albrecht:1999zu},
in radiative $J/\psi$ decay \cite{Mallik:1989if,Eigen:1990cm},
in $K^{-} p \to K^{*} \bar{K}^{*} \Lambda$ reaction 
\cite{Aston:1989gx,Aston:1990wf}, 
and in central production in proton-proton collisions 
\cite{Armstrong:1986cm,Armstrong:1989es,Barberis:1998tv}.
It is known from the WA102 experiment \cite{Barberis:1998tv} that
although the $K^{*0} \bar{K}^{*0}$ final state is a major component
of the $K^{+}K^{-}\pi^{+}\pi^{-}$ channel it is not the dominant component. 
In contrast, the $\phi \phi$ final state was found to be
dominant component of the $K^{+}K^{-}K^{+}K^{-}$ channel \cite{Barberis:1998bq}.
The cross section as a function of center-of-mass energy 
for the production of $K^{*}(892) \bar{K}^{*}(892)$ system
was found \cite{Barberis:1998tv}
to be consistent with being produced via the double-pomeron-exchange mechanism.

In hadronic proton-proton collisions
\cite{Armstrong:1989es,Barberis:1998tv,Breakstone:1989ty}
a broad low-mass enhancement in the $K^{*} \bar{K}^{*}$ and/or 
$K^{+}K^{-}\pi^{+}\pi^{-}$ invariant mass distributions was seen.
In \cite{Armstrong:1989es} the authors stated that 
the $K^{*0} \bar{K}^{*0}$ system is mainly produced
as a nonresonant threshold enhancement.
More recent analysis \cite{Barberis:2000em}
give some evidence for the existence of $f_{2}(1950)$ resonance 
in the $K^{*0} \bar{K}^{*0}$ channel; 
see Fig.~3~(c) of \cite{Barberis:2000em}.
On the other hand, 
in the radiative $J/\psi$ decay \cite{Mallik:1989if,Eigen:1990cm}
the $K^{*} \bar{K}^{*}$ spectrum indicates two narrow peaks at low mass.
The analysis of angular distributions finds
that the $K^{*} \bar{K}^{*}$ system in the radiative $J/\psi$ decay
show strong $J^{PC} = 0^{-+}$ component whereas the hadronic production modes 
are all consistent with strong (broad) $J^{PC} = 2^{++}$ component.
The analysis of the partial wave structure of the $K^{*0} \bar{K}^{*0}$ state 
from the reaction $\gamma \gamma \to K^{+}K^{-}\pi^{+}\pi^{-}$
\cite{Albrecht:1999zu} support the dominance 
of the $(J^{P},J_{z})=(2^{+},\pm 2)$ wave.

An interesting suggestion has been made for the broad isoscalar-tensor $f_{2}(1950)$ resonance
to be the lightest tensor glueball,
while the arguments are not yet fully settled.
Namely, this state is occasionally discussed as a candidate for a tensor glueball 
as it appears to have largely flavor-blind decay modes;
see e.g. \cite{Brunner:2015oqa,Godizov:2016vuw,Zhang:2016vcx}.
However, according to lattice-QCD simulations, the lightest tensor glueball 
has a mass between 2.2~GeV and 2.4~GeV, see, e.g., 
\cite{Morningstar:1999rf,Gregory:2012hu,Sun:2017ipk}.
Thus, the $f_{2}(2300)$ and $f_{2}(2340)$ states 
are good candidates to be tensor glueball.
The nature of these resonances is not understood at present
and a tensor glueball has still not been clearly identified.
Nontrivial are predictions of not only masses
but also widths of the predicted glueballs;
see, e.g., Sec.~8 of \cite{Szanyi:2019kkn}
for more information on this topic.
Glueballs are expected to lie on the pomeron trajectory.
It was shown in \cite{Szanyi:2019kkn} 
that even small variations of the parameters in the 
(nonlinear) pomeron/glueball trajectory 
result in noticeable changes of glueballs widths.

It is also interesting to speculate whether 
the tensor states $f_{2}(1910)$, $f_{2}(1950)$, and $f_{2}(2150)$,
observed by the WA102 Collaboration \cite{Kirk:2000ws},
are due to mixing between a tensor glueball 
and nearby $q \bar{q}$ states.
Two of these states have similar $\phi_{pp}$ and $\rm{dP_{t}}$ 
dependencies and one the opposite;
$\phi_{pp}$ is the azimuthal angle between the transverse momentum vectors 
of the outgoing protons, 
and $\rm{dP_{t}}$ is the so-called ``glueball-filter variable'' \cite{Close:1997pj}
defined by the difference of the transverse momentum vectors of the outgoing protons.
It is known from the WA102 analysis of various channels 
that all the undisputed $q \bar{q}$ states 
are suppressed at small $\rm{dP_{t}}$ in contrast to glueball candidates.
Established $q \bar{q}$ states peak at $\phi_{pp} = \pi$
whereas the $f_{2}(1910)$ and $f_{2}(1950)$ peak 
at $\phi_{pp} = 0$ \cite{Kirk:2000ws}.
These experimental observations in central production
indicate that the $f_{2}(1950)$ probably
contains large gluonic component and should be copiously produced 
via the double-pomeron-exchange (i.e., $\Pom \Pom$-fusion) mechanism.

However, the observation of $f_{2}(1950)$ resonance  
in two-photon interaction processes,
such as $\gamma \gamma \to f_{2}(1950) \to K^{*0} \bar{K}^{*0}$
\cite{Albrecht:1987jp,Albrecht:1988gr,Albrecht:1999zu},
and in other $\gamma \gamma$-fusion processes
\cite{Abe:2003vn,Uehara:2009cka,Klusek-Gawenda:2017lgt},
precludes its interpretation as a pure gluonic state.
In \cite{Klusek-Gawenda:2017lgt} 
a good description the Belle data on $\gamma \gamma \to p \bar{p}$
including, in addition to the proton exchange, 
the $f_{2}(1950)$ resonance was obtained.
One can observe there the dominance of the $f_{2}(1950)$ resonance
in the low mass region $M_{p \bar{p}} = W_{\gamma \gamma} < 2.5$~GeV.
In the $f_{2}(1950)$-exchange amplitude only the term 
with $a_{f_{2}(1950) \gamma \gamma}$ coupling
and $g^{(2)}_{f_{2}(1950) p \bar{p}}$ coupling was used.
There, $a$- and $b$-type coupling parametrise 
the so-called helicity-zero and helicity-two 
$\gamma \gamma \to f_{2}$ amplitudes, respectively;
see e.g. \cite{Ewerz:2013kda}.
For instance, 
for the $\gamma \gamma \to f_{2}(1270)$ process
the helicity-2 contribution ($b$-type coupling) is dominant.
As will be presented in this work, 
for diffractive processes shown in Fig.~\ref{fig:diagrams},
the $b$-type coupling in the $f_{2}(1950) K^{*0} \bar{K}^{*0}$ 
and $\Pom K^{*} K^{*}$ vertices
is more preferred than the $a$-type coupling.

The study of $\phi \phi$ and $K^{*} \bar{K}^{*}$ systems
could provide also helpful information for searching
for the fully-strange ($ss \bar{s} \bar{s}$) tetraquark.
In the relativistic quark model 
based on the quasipotential approach in QCD \cite{Ebert:2008id},
the $f_{2}(1950)$ and $f_{2}(2340)$ states
are considered as a candidates for the ground state 
($\langle \textbf{L}^{2} \rangle = 0$) 
light tetraquarks as diquark-antidiquark 
(composed from an axial vector diquark and antidiquark),
${qq} {\bar{q}\bar{q}}$ and ${ss} {\bar{s}\bar{s}}$, respectively.
In \cite{Lu:2019ira} 
the $f_{2}(2300)$ is assigned 
to be $ss \bar{s} \bar{s}$ tetraquark state.
These two states $f_{2}(2300)$ and $f_{2}(2340)$
are close in mass within errors \cite{Zyla:2020zbs}.
Very recently, in \cite{Liu:2020lpw} it was stated 
that the $f_{2}(2340)$ resonance
may be assigned to $1S$-wave tetraquark $T_{ss \bar{s}\bar{s}}(2381)$ 
in the framework of a nonrelativistic potential
quark model without the diquark-antidiquark approximation.
The $f_{2}(2340)$ state may have large decay rates 
into the $\phi \phi$ and $\eta \eta$ final states
through quark rearrangements, 
and/or into $K^{*} \bar{K}^{*}$ final state 
through the annihilation of $s \bar{s}$ 
and creation of a pair of nonstrange $q \bar{q}$.
To confirm this assignment, the above decay modes
and such as $\eta \eta'$, $\eta' \eta'$
should be investigated in experimental searches.
On the other hand, flavor mixings could be important 
for the light flavor systems and pure $ss \bar{s} \bar{s}$ states 
may not exist; see, e.g., \cite{Kim:2018zob}.

With the idea of bringing more information on the topic, 
in the present work, we study
the diffractive $\Pom \Pom \to f_{2}(1950)$ fusion mechanism
in the reaction $pp \to pp K^{*0} \bar{K}^{*0}$
and the $K^{*0} \bar{K}^{*0}$ continuum
which is a background for diffractively produced resonances.
But we emphasize that in the following we make no assumptions
on whether the $f_{2}(1950)$ resonance is glueball or tetraquark.
The problem is interesting because of the production 
of expected glueballs can be associated with other mesons 
and can be accompanied by a diffractive continuum 
making the identification of glueballs 
rather difficult.

In the tensor-pomeron model for soft high-energy scattering
formulated in \cite{Ewerz:2013kda},
on the basis of earlier work \cite{Nachtmann:1991ua},
the pomeron exchange is effectively treated 
as the exchange of a rank-2 symmetric tensor.
In the last few years a scientific program was undertaken to analyse 
the central exclusive production of mesons 
in the tensor-pomeron model in several reactions: 
$p p \to p p M$~\cite{Lebiedowicz:2013ika},
where $M$ stands for a scalar or pseudoscalar meson, 
$p p \to p p \pi^{+}\pi^{-}$ and $pp \to p p (f_{2}(1270) \to \pi^{+}\pi^{-})$
\cite{Lebiedowicz:2014bea,Lebiedowicz:2016ioh,Lebiedowicz:2019por}, 
$p p \to p n \rho^{0} \pi^{+}$ ($p p \rho^{0} \pi^{0}$) 
\cite{Lebiedowicz:2016ryp},
$p p \to p p K^{+}K^{-}$ \cite{Lebiedowicz:2018eui},
$p p \to p p (\sigma \sigma, \rho^{0} \rho^{0} \to \pi^{+}\pi^{-}\pi^{+}\pi^{-})$ 
\cite{Lebiedowicz:2016zka},
$p p \to p p p \bar{p}$ \cite{Lebiedowicz:2018sdt},
$p p \to p p (\phi \phi \to K^{+}K^{-}K^{+}K^{-})$ 
\cite{Lebiedowicz:2019jru},
$p p \to p p (\phi \to K^{+}K^{-}, \mu^{+}\mu^{-})$ 
\cite{Lebiedowicz:2019boz},
$p p \to p p f_{1}(1285)$ and $p p \to p p f_{1}(1420)$ 
\cite{Lebiedowicz:2020yre}. 
The present paper aims to underline the importance of the study 
of the $pp \to pp(K^{*0} \bar{K}^{*0} \to K^{+} \pi^{-} K^{-} \pi^{+})$ reaction.

Some effort to measure exclusive
production of higher-multiplicity central systems 
at the energy $\sqrt{s} = 13$~TeV 
has been initiated by the ATLAS Collaboration; 
see, e.g., \cite{Sikora:2747846}.
We think that a study of CEP of the $K^{*0} \bar{K}^{*0}$ pairs
decaying into $K^{+} \pi^{-} K^{-} \pi^{+}$
should be quite rewarding for experimentalists.
Our analysis are designed to facilitate 
the study of such processes at the LHC,
for instance, by investigating in detail 
the continuum and tensor resonance production.

The paper is organized as follows. 
In Sec.~\ref{section:formalism} we describe our theoretical framework.
In Sec.~\ref{sec:results} we show and discuss our numerical results.
We determine the model parameters from a comparison 
to the WA102 experimental data for the reaction 
$pp \to pp K^{*0} \bar{K}^{*0}$.
We also predict the total and differential cross sections for 
$pp \to pp (K^{*0} \bar{K}^{*0} \to K^{+} K^{-} \pi^{+} \pi^{-})$
including typical kinematic cuts for the LHC experiments.
The final section is devoted to the conclusions.

\section{Theoretical framework}
\label{section:formalism}

\begin{figure}
(a)\includegraphics[width=7.5cm]{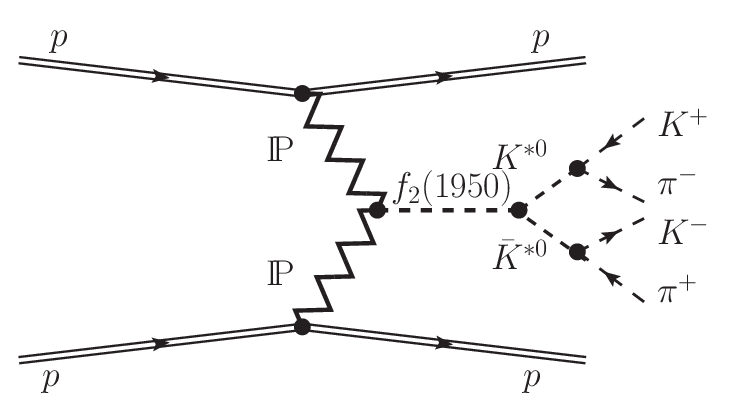}  
(b)\includegraphics[width=6.5cm]{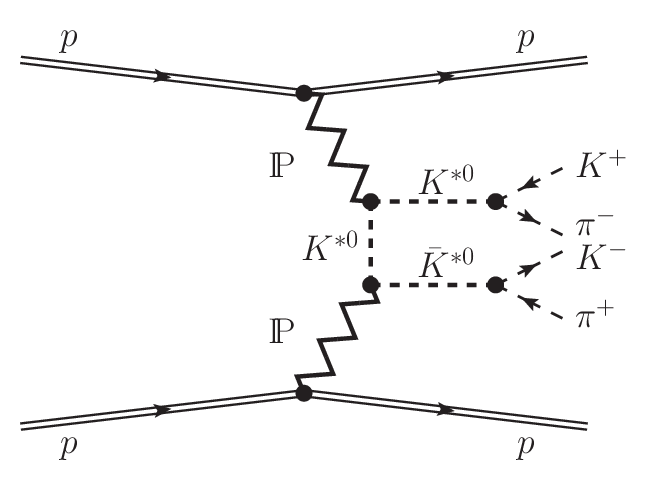}    
\caption{\label{fig:diagrams}
The ``Born level'' diagrams for double pomeron
central exclusive $K^{*0} \bar{K}^{*0}$ production 
and their subsequent decays into $K^{+}\pi^{-}K^{-}\pi^{+}$ 
in proton-proton collisions:
(a) $K^{*0} \bar{K}^{*0}$ production via the $f_{2}(1950)$ resonance;
(b) continuum $K^{*0} \bar{K}^{*0}$ production.}
\end{figure}

In the present paper we consider two processes shown in Fig.~\ref{fig:diagrams}
that may contribute to the $K^{+}\pi^{-}K^{-}\pi^{+}$
final state 
via an intermediate $K^{*0} \bar{K}^{*0}\equiv K^{*0}(892) \bar{K}^{*0}(892)$.
Figure~\ref{fig:diagrams}(a) shows the process with intermediate production
of $f_{2}(1950)$ resonance,
\begin{eqnarray}
pp \to pp \, (\Pom \Pom \to f_{2}(1950) \to K^{*0} \bar{K}^{*0}) \to pp \,K^{+} \pi^{-} K^{-} \pi^{+}\,.
\label{2to6_reaction_f2}
\end{eqnarray} 
In Fig.~\ref{fig:diagrams}(b) we have the continuum process
\begin{eqnarray}
pp \to pp \, (\Pom \Pom \to K^{*0} \bar{K}^{*0}) \to pp \,K^{+} \pi^{-} K^{-} \pi^{+}\,
\label{2to6_reaction_continuum}
\end{eqnarray} 
with the $K^{*0}(892)$ $t/u$-channel exchanges.
%

The processes (\ref{2to6_reaction_f2})
and (\ref{2to6_reaction_continuum}) 
are expected to be most important ones
at high energies since they involve pomeron exchange only.
We can replace one or two pomerons
by one or two $f_{2 \Reg}$ reggeons.
However, for the LHC collision energies and central 
$K^{*0} \bar{K}^{*0}$ production (midrapidity region)
such $f_{2 \Reg} f_{2 \Reg}$-, $f_{2 \Reg} \Pom$-, 
and $\Pom f_{2 \Reg}$-fusion contributions 
are expected to be small 
and we shall not consider them in our present paper.

We treat effectively the $2 \to 6$ processes (\ref{2to6_reaction_f2})
and (\ref{2to6_reaction_continuum}) as arising 
from the $pp \to pp  K^{*0} \bar{K}^{*0}$ reaction.
The general cross-section formula can be written approximately as
\begin{eqnarray}
{\sigma}_{2 \to 6} =
\int_{m_{K}+m_{\pi}}^{{\rm max}\{m_{X_{3}}\}} 
\int_{m_{K}+m_{\pi}}^{{\rm max}\{m_{X_{4}}\}}
{\sigma}_{2 \to 4}(...,m_{X_{3}},m_{X_{4}})
f_{K^{*}}(m_{X_{3}}) \,  
f_{K^{*}}(m_{X_{4}}) \, dm_{X_{3}} \, dm_{X_{4}}\,.
\label{2to6_amplitude}
\end{eqnarray}
We use for the calculation of decay processes $K^{*} \to K \pi$
the spectral function
\begin{eqnarray}
f_{K^{*}}(m_{X_{i}}) = C_{K^{*}}\,
\left( 1-\dfrac{(m_{K}+m_{\pi})^{2}}{m_{X_{i}}^{2}} \right)^{3/2}
\frac{\frac{2}{\pi}{m_{X_{i}}m_{K^{*}}} \Gamma_{K^{*}}}{(m_{X_{i}}^{2}-m_{K^{*}}^{2})^{2} + m_{K^{*}}^{2} \Gamma_{K^{*}}^{2}}\,,
\label{spectral_function}
\end{eqnarray}
where $i  = 3, 4$, $\Gamma_{K^{*}}$ is the total width of the $K^{*}(892)$ resonance and
$m_{K^{*}}$ its mass taken from \cite{Zyla:2020zbs},
the factor $C_{K^{*}}$ is found from the condition
\begin{eqnarray}
\int_{m_{K} + m_{\pi}}^{{\rm max}\{m_{X}\}} f_{K^{*}}(m_{X}) \,dm_{X} = 1\,.
\label{spectral_function_aux}
\end{eqnarray}

To include experimental cuts on outgoing pseudoscalar
particles we perform the decays of $K^{*}(892)$ mesons
isotropically in the $K^{*}$-meson rest frame and then use
relativistic transformations to the overall center-of-mass frame.

Now we discuss the production of $K^{*0} \bar{K}^{*0}$ 
in proton-proton collisions,
\begin{eqnarray}
p(p_{a},\lambda_{a}) + p(p_{b},\lambda_{b}) \to
p(p_{1},\lambda_{1}) + p(p_{2},\lambda_{2}) + 
K^{*0}(p_{3},\lambda_{3}) + \bar{K}^{*0}(p_{4},\lambda_{4}) \,,
\label{2to4_reaction}
\end{eqnarray}
where 
$p_{a,b}$, $p_{1,2}$ and $\lambda_{a,b}, \lambda_{1,2} = \pm \frac{1}{2}$ 
denote the four-momenta and helicities of the protons and
$p_{3,4}$ and $\lambda_{3,4} = 0, \pm 1$ 
denote the four-momenta and helicities of the $K^{*0}$ mesons, respectively.

The amplitude for the $2 \to 4$ reaction (\ref{2to4_reaction})
can be written as
\begin{eqnarray}
\begin{split}
{\cal M}_{\lambda_{a}\lambda_{b}\to\lambda_{1}\lambda_{2} K^{*} \bar{K}^{*}} = 
\left(\epsilon^{(K^{*})}_{\kappa_{3}}(\lambda_{3})\right)^*
\left(\epsilon^{(\bar{K}^{*})}_{\kappa_{4}}(\lambda_{4})\right)^*
{\cal M}^{\kappa_{3}\kappa_{4}}_{\lambda_{a}\lambda_{b}\to\lambda_{1}\lambda_{2} K^{*} \bar{K}^{*}}\,,
\end{split}
\label{amplitude_2to4}
\end{eqnarray}
where $\epsilon^{(K^{*})}_{\kappa}$ 
are the polarisation vectors of the $K^{*}$ mesons.
Taking into account summation over the $K^{*}$ polarisations
we get for the amplitudes squared 
[to be inserted in $\sigma_{2 \to 4}$ in Eq.~(\ref{2to6_amplitude})]
\begin{eqnarray}
\frac{1}{4} \sum_{\rm{spins}}
\Big|{\cal M}_{\lambda_{a}\lambda_{b}\to\lambda_{1}\lambda_{2} K^{*} \bar{K}^{*}}\Big|^{2}
=\frac{1}{4} \sum_{\lambda_{a},\lambda_{b},\lambda_{1},\lambda_{2}}
\left({\cal M}^{\sigma_{3} \sigma_{4}}_{\lambda_{a}\lambda_{b}\to\lambda_{1}\lambda_{2} K^{*} \bar{K}^{*}}\right)^{*}
{\cal M}^{\rho_{3} \rho_{4}}_{\lambda_{a}\lambda_{b}\to\lambda_{1}\lambda_{2} K^{*} \bar{K}^{*}}\,
g_{\sigma_{3}\rho_{3}} \,g_{\sigma_{4}\rho_{4}}
\,.\nonumber \\
\label{amplitude_squared_2to4}
\end{eqnarray}

We take into account two main processes
shown by the diagrams in Fig.~\ref{fig:diagrams}.
The full amplitude is then the sum of 
the $f_{2}(1950)$ resonance term
and the $K^{*0}$-exchange continuum term:
\begin{eqnarray}
\begin{split}
{\cal M}^{\kappa_{3}\kappa_{4}}_{\lambda_{a}\lambda_{b}\to\lambda_{1}\lambda_{2} K^{*} \bar{K}^{*}}
= {\cal M}^{(\Pom \Pom \to f_{2} \to K^{*}\bar{K}^{*})\,
\kappa_{3} \kappa_{4}}_{\lambda_{a}\lambda_{b}
\to\lambda_{1}\lambda_{2} K^{*}\bar{K}^{*}}
+ {\cal M}^{(K^{*}\mathrm{-exchange})\,
\kappa_{3} \kappa_{4}}_{\lambda_{a}\lambda_{b}\to\lambda_{1}\lambda_{2} K^{*} \bar{K}^{*}}
\,.
\end{split}
\label{amplitude_sum}
\end{eqnarray}
In our exploratory study we consider these terms separately, i.e.,
we neglect interference effects between 
the resonant $f_{2}(1950) \to K^{*0} \bar{K}^{*0}$ 
and the continuum $K^{*0} \bar{K}^{*0}$ processes. 

To give the full physical amplitude for the reaction (\ref{2to4_reaction})
we include absorptive corrections to the Born amplitudes 
in the one-channel eikonal approximation;
see, e.g., Sec.~3.3 of \cite{Lebiedowicz:2014bea}.
In practice we work with the amplitudes in the high-energy approximation,
i.e., assume $s$-channel helicity conservation for the protons.

\subsection{$f_{2}(1950)$ resonance contribution}
\label{sec:f2_KKbar}

Now we consider the amplitude representing 
by the diagram in Fig.~\ref{fig:diagrams}(a)
but limiting to the final state $pp K^{*0} \bar{K}^{*0}$.

The Born-level amplitude 
for the $\Pom \Pom$-fusion process 
through the $s$-channel $f_{2}(1950)$-meson exchange
is given by
\begin{eqnarray}
{\cal M}^{(\Pom \Pom \to f_{2} \to K^{*}\bar{K}^{*})\,\kappa_{3} \kappa_{4}}_{\lambda_{a}\lambda_{b}
\to\lambda_{1}\lambda_{2} K^{*}\bar{K}^{*}} 
= && (-i)\,
\bar{u}(p_{1}, \lambda_{1}) 
i\Gamma^{(\Pom pp)\,\mu_{1} \nu_{1}}(p_{1},p_{a}) 
u(p_{a}, \lambda_{a})\;
i\Delta^{(\Pom)}_{\mu_{1} \nu_{1}, \alpha_{1} \beta_{1}}(s_{1},t_{1}) \nonumber \\
&& \times 
i\Gamma^{(\Pom \Pom f_{2})\,\alpha_{1} \beta_{1},\alpha_{2} \beta_{2}, \rho \sigma}(q_{1},q_{2}) \;
i\Delta^{(f_{2})}_{\rho \sigma, \alpha \beta}(p_{34})\;
i\Gamma^{(f_{2} K^{*}\bar{K}^{*})\,\alpha \beta \kappa_{3} \kappa_{4}}(p_{3},p_{4}) \nonumber \\
&& \times 
i\Delta^{(\Pom)}_{\alpha_{2} \beta_{2}, \mu_{2} \nu_{2}}(s_{2},t_{2}) \;
\bar{u}(p_{2}, \lambda_{2}) 
i\Gamma^{(\Pom pp)\,\mu_{2} \nu_{2}}(p_{2},p_{b}) 
u(p_{b}, \lambda_{b}) \,, \nonumber \\
\label{amplitude_f2_pomTpomT}
\end{eqnarray}
where
$s_{1} = (p_{1} + p_{3} + p_{4})^{2}$,
$s_{2} = (p_{2} + p_{3} + p_{4})^{2}$, 
$q_{1} = p_{a} - p_{1}$, 
$q_{2} = p_{b} - p_{2}$, 
$t_{1} = q_{1}^{2}$, $t_{2} = q_{2}^{2}$,
and
$p_{34} = q_{1} + q_{2} = p_{3} + p_{4}$.
Here $\Gamma^{(\Pom pp)}$ and $\Delta^{(\Pom)}$
denote the effective proton vertex function and propagator, respectively, for the tensor-pomeron exchange.
The corresponding expressions, as given in Sec.~3 of \cite{Ewerz:2013kda}, are as follows
\begin{eqnarray}
&&i\Gamma_{\mu \nu}^{(\Pom pp)}(p',p)
=-i 3 \beta_{\Pom NN} F_{1}(t)
\left\lbrace 
\frac{1}{2} 
\left[ \gamma_{\mu}(p'+p)_{\nu} 
     + \gamma_{\nu}(p'+p)_{\mu} \right]
- \frac{1}{4} g_{\mu \nu} ( p\!\!\!/' + p\!\!\!/ )
\right\rbrace, \qquad \;\;
\label{Ppp_vertex}\\
&&i \Delta^{(\Pom)}_{\mu \nu, \kappa \lambda}(s,t) = 
\frac{1}{4s} \left( g_{\mu \kappa} g_{\nu \lambda} 
                  + g_{\mu \lambda} g_{\nu \kappa}
                  - \frac{1}{2} g_{\mu \nu} g_{\kappa \lambda} \right)
(-i s \alpha'_{\Pom})^{\alpha_{\Pom}(t)-1}\,,
\label{P_propagator}
\end{eqnarray}
where $\beta_{\Pom NN} =1.87$~GeV$^{-1}$ and 
$F_{1}(t)$ is the Dirac form factor of the proton.
For extensive discussions of the properties of these terms 
we refer to \cite{Ewerz:2013kda}.
In (\ref{P_propagator}) the pomeron trajectory $\alpha_{\Pom}(t)$
is assumed to be of standard linear form, see e.g. \cite{Donnachie:1992ny,Donnachie:2002en},
\begin{eqnarray}
&&\alpha_{\Pom}(t) = \alpha_{\Pom}(0)+\alpha'_{\Pom}\,t, \nonumber \\
&&\alpha_{\Pom}(0) = 1.0808\,, \;\;
\alpha'_{\Pom} = 0.25 \; {\rm GeV}^{-2}\,.
\label{trajectory}
\end{eqnarray}

The $\Pom \Pom f_{2}$ vertex, including a form factor,
can be written as
\begin{eqnarray}
i\Gamma_{\mu \nu,\kappa \lambda,\rho \sigma}^{(\Pom \Pom f_{2})} (q_{1},q_{2}) =
\left( i\Gamma_{\mu \nu,\kappa \lambda,\rho \sigma}^{(\Pom \Pom f_{2})(1)} \mid_{\rm bare}
+ \sum_{j=2}^{7}i\Gamma_{\mu \nu,\kappa \lambda,\rho \sigma}^{(\Pom \Pom f_{2})(j)}(q_{1},q_{2}) \mid_{\rm bare} 
\right)
\tilde{F}^{(\Pom \Pom f_{2})}(q_{1}^{2},q_{2}^{2},p_{34}^{2}) \,.\nonumber\\
\label{vertex_pompomT}
\end{eqnarray}
A possible choice for the 
$i\Gamma_{\mu \nu,\kappa \lambda,\rho \sigma}^{(\Pom \Pom f_{2})(j)}$
coupling terms $j = 1, ..., 7$ 
is given in Appendix~A of \cite{Lebiedowicz:2016ioh}.
The couplings $j = 1, ..., 7$ can be associate to the following
orbital angular momentum and spin of the two ``real pomerons''
$(l,S)$ values:
$(0,2)$, $(2,0)-(2,2)$, $(2,0)+(2,2)$, $(2,4)$, 
$(4,2)$, $(4,4)$, $(6,4)$, respectively.
In the following we shall, for the purpose of orientation,
assume that only the $j= 1$ coupling in (\ref{vertex_pompomT}),
corresponding to the lowest values of $(l,S)$, that is $(l,S) = (0,2)$,
is unequal to zero.
The expressions for $j= 1$ vertex is as follows:\footnote{Here 
the label ``bare'' is used
for a vertex, as derived from a corresponding coupling Lagrangian
in Appendix~A of \cite{Lebiedowicz:2016ioh} without a form-factor function.}
\begin{eqnarray}
&&i\Gamma_{\mu \nu,\kappa \lambda,\rho \sigma}^{(\Pom \Pom f_{2})(1)} \mid_{\rm bare}
=
2 i \,g^{(1)}_{\Pom \Pom f_{2}} M_{0}\, 
R_{\mu \nu \mu_{1} \nu_{1}}\,
R_{\kappa \lambda \alpha_{1} \lambda_{1}}\,
R_{\rho \sigma \rho_{1} \sigma_{1}}\,
g^{\nu_{1} \alpha_{1}}\,g^{\lambda_{1} \rho_{1}}\,g^{\sigma_{1} \mu_{1}}\,, 
\label{A1}\\
&&R_{\mu \nu \kappa \lambda} = \frac{1}{2} g_{\mu \kappa} g_{\nu \lambda} 
                           + \frac{1}{2} g_{\mu \lambda} g_{\nu \kappa}
                            -\frac{1}{4} g_{\mu \nu} g_{\kappa \lambda}\,,
\label{A2}
\end{eqnarray}
see (A12) of \cite{Lebiedowicz:2016ioh}. In (\ref{A1}),
$M_{0} \equiv 1$~GeV and the $g^{(1)}_{\Pom \Pom f_{2}}$ 
is dimensionless coupling constant 
which should be fitted to experimental data.
We take the factorized form for the $\Pom \Pom f_{2}$ form factor
in (\ref{vertex_pompomT}):
\begin{eqnarray}
&&\tilde{F}^{(\Pom \Pom f_{2})}(q_{1}^{2},q_{2}^{2},p_{34}^{2}) = 
\tilde{F}_{M}(q_{1}^{2}) \tilde{F}_{M}(q_{2}^{2}) F^{(\Pom \Pom f_{2})}(p_{34}^{2})\,;
\label{Fpompommeson}\\
&&\tilde{F}_{M}(t)=\frac{1}{1-t/\tilde{\Lambda}_{0}^{2}}\,.
\label{FM_t}
\end{eqnarray}
The form factor $F^{(\Pom \Pom f_{2})}$
is normalized to unity at the on-shell point
$F^{(\Pom \Pom f_{2})}(m_{f_{2}}^{2}) = 1$
and parametrised in two ways:
\begin{eqnarray}
&&F^{(\Pom \Pom f_{2})}(p_{34}^{2}) = 
\exp{ \left( \frac{-(p_{34}^{2}-m_{f_{2}}^{2})^{2}}{\Lambda_{f_{2},E}^{4}} \right)}\,,
\label{Fpompommeson_ff_E} \\
&&F^{(\Pom \Pom f_{2})}(p_{34}^{2}) = 
\frac{\Lambda_{f_{2},P}^4}
{\Lambda_{f_{2},P}^4 + (p_{34}^{2} - m_{f_{2}}^2)^{2}}\,.
\label{Fpompommeson_ff_P}
\end{eqnarray}
The cutoff parameters
$\tilde{\Lambda}_{0}$,
$\Lambda_{f_{2},E}$ and $\Lambda_{f_{2},P}$ 
in (\ref{FM_t}), (\ref{Fpompommeson_ff_E}) and (\ref{Fpompommeson_ff_P}),
respectively,
are treated as free parameters
which could be adjusted to fit the experimental data.

We use in (\ref{amplitude_f2_pomTpomT}) the tensor-meson propagator 
with the simple Breit-Wigner form; see (3.35) of \cite{Lebiedowicz:2019jru}.
A better representation for the propagator could be constructed using
the methods of \cite{Melikhov:2003hs,Ewerz:2013kda}, 
used there for the $\rho^{0}$ and $f_{2}(1270)$ propagators.
In our calculations we take the nominal values 
for the $f_{2}(1950)$ from \cite{Zyla:2020zbs}:
\begin{eqnarray}
m_{f_{2}} &=& (1936 \pm 12)\;\,{\rm MeV}\,, \nonumber \\
\Gamma_{f_{2}} &=& (464 \pm 24) \;\,{\rm MeV}\,.
\label{f2_parameters}
\end{eqnarray}

For the $f_{2} K^{*}\bar{K}^{*}$ vertex function we take the same ansatz
as for the $f_{2} \phi \phi$ vertex defined in (3.32) of \cite{Lebiedowicz:2019jru}.
The $f_{2} K^{*}\bar{K}^{*}$ vertex is as follows:
\begin{eqnarray}
i\Gamma^{(f_{2} K^{*}\bar{K}^{*})}_{\mu \nu \kappa \lambda}(p_{3},p_{4}) &=&
i\dfrac{2}{M_{0}^{3}} g'_{f_{2} K^{*}\bar{K}^{*}}\,  
\Gamma^{(0)}_{\mu \nu \kappa \lambda}(p_{3},p_{4})\,
F'^{(f_{2} K^{*}\bar{K}^{*})}(p_{34}^{2}) \nonumber \\
&&- i \dfrac{1}{M_{0}} g''_{f_{2} K^{*}\bar{K}^{*}}\,\Gamma^{(2)}_{\mu \nu \kappa \lambda}(p_{3},p_{4})\,
F''^{(f_{2} K^{*}\bar{K}^{*})}(p_{34}^{2})
\label{vertex_f2KKbar}
\end{eqnarray}  
with two rank-four tensor functions,
\begin{eqnarray}
&&\Gamma_{\mu\nu\kappa\lambda}^{(0)} (p_3,p_4) =
\Big[(p_3 \cdot p_4) g_{\mu\nu} - p_{4\mu} p_{3\nu}\Big] 
\Big[p_{3\kappa}p_{4\lambda} + p_{4\kappa}p_{3\lambda} - 
\frac{1}{2} (p_3 \cdot p_4) g_{\kappa\lambda}\Big] \,,
\label{3.16}\\
&&\Gamma_{\mu\nu\kappa\lambda}^{(2)} (p_3,p_4) = \,
 (p_3\cdot p_4) (g_{\mu\kappa} g_{\nu\lambda} + g_{\mu\lambda} g_{\nu\kappa} )
+ g_{\mu\nu} (p_{3\kappa} p_{4\lambda} + p_{4\kappa} p_{3\lambda} ) \nonumber \\
&& \qquad \qquad \qquad \quad - p_{3\nu} p_{4 \lambda} g_{\mu\kappa} - p_{3\nu} p_{4 \kappa} g_{\mu\lambda} 
- p_{4\mu} p_{3 \lambda} g_{\nu\kappa} - p_{4\mu} p_{3 \kappa} g_{\nu\lambda} 
\nonumber \\
&& \qquad \qquad \qquad \quad - [(p_3 \cdot p_4) g_{\mu\nu} - p_{4\mu} p_{3\nu} ] \,g_{\kappa\lambda} \,;
\label{3.17}
\end{eqnarray}
see Eqs.~(3.18) and (3.19) of \cite{Ewerz:2013kda}.
The coupling parameters
$g'_{f_{2} K^{*}\bar{K}^{*}}$ and $g''_{f_{2} K^{*}\bar{K}^{*}}$ 
are dimensionless.
Different form factors $F'$ and $F''$ are allowed
\textit{a~priori}.
We assume, in the present exploratory study, that
\begin{eqnarray}
F'^{(f_{2} K^{*}\bar{K}^{*})}(p_{34}^{2}) = 
F''^{(f_{2} K^{*}\bar{K}^{*})}(p_{34}^{2}) = 
F^{(\Pom \Pom f_{2})}(p_{34}^{2})
\label{Fpompommeson_ff_tensor}
\end{eqnarray}
and for the cutoff parameters to be the same,
$\Lambda'_{f_{2}} = \Lambda''_{f_{2}} = \Lambda_{f_{2},E}$ 
or $\Lambda_{f_{2},P}$;
see (\ref{Fpompommeson_ff_E}) and (\ref{Fpompommeson_ff_P}).

One has to keep in mind that relative signs of couplings 
have physical significance, for instance, the relative sign of $g'$ and $g''$.
However, for orientation purposes, in the calculation 
we treat them separately and do not fix the sign of the $f_{2}$ couplings.
With our choice to keep only one $\Pom \Pom f_{2}(1950)$ coupling
from (\ref{vertex_pompomT}), namely (\ref{A1}) with $g^{(1)}_{\Pom \Pom f_{2}}$,
the results will depend on the product of the couplings
$g^{(1)}_{\Pom \Pom f_{2}} g'_{f_{2} K^{*}\bar{K}^{*}}$ and
$g^{(1)}_{\Pom \Pom f_{2}} g''_{f_{2} K^{*}\bar{K}^{*}}$
with $g'_{f_{2} K^{*}\bar{K}^{*}}$ and $g''_{f_{2} K^{*}\bar{K}^{*}}$
given in (\ref{vertex_f2KKbar}).
In the following we assume that only either
the first or the second of the above products of couplings 
is nonzero.

\subsection{Diffractive continuum contribution} 
\label{sec:pompom_KKbar}

The diagram for the continuum $K^{*0} \bar{K}^{*0}$
with an intermediate $K^{*0}$ exchange
is shown in Fig.~\ref{fig:diagrams}~(b).
The Born-level amplitude can be written as the sum
\begin{eqnarray}
{\cal M}^{(K^{*}\mathrm{-exchange})\,\kappa_{3} \kappa_{4}}_{\lambda_{a}\lambda_{b}\to\lambda_{1}\lambda_{2} K^{*} \bar{K}^{*}} =
{\cal M}^{({\hat{t}})\,\kappa_{3} \kappa_{4}}_{\lambda_{a} \lambda_{b} \to \lambda_{1} \lambda_{2} 
K^{*} \bar{K}^{*}} +
{\cal M}^{({\hat{u}})\,\kappa_{3} \kappa_{4}}_{\lambda_{a} \lambda_{b} \to \lambda_{1} \lambda_{2} 
K^{*} \bar{K}^{*}} 
\label{amp_continuum}
\end{eqnarray}
with the $\hat{t}$- and $\hat{u}$-channel amplitudes:
\begin{eqnarray}
{\cal M}^{({\hat{t}})}_{\kappa_{3} \kappa_{4}} = 
&& (-i) \bar{u}(p_{1}, \lambda_{1}) 
i\Gamma^{(\Pom pp)}_{\mu_{1} \nu_{1}}(p_{1},p_{a}) 
u(p_{a}, \lambda_{a})\,
i\Delta^{(\Pom)\, \mu_{1} \nu_{1}, \alpha_{1} \beta_{1}}(s_{13},t_{1})  \nonumber \\
&& \times 
i\Gamma^{(\Pom K^{*}K^{*})}_{\kappa_{1} \kappa_{3} \alpha_{1} \beta_{1}}(\hat{p}_{t},-p_{3})\,
i\Delta^{(K^{*})\,\kappa_{1} \kappa_{2}}(\hat{p}_{t})\,
i\Gamma^{(\Pom K^{*}K^{*})}_{\kappa_{4} \kappa_{2} \alpha_{2} \beta_{2}}(p_{4},\hat{p}_{t})  \nonumber \\
&& \times 
i\Delta^{(\Pom)\, \alpha_{2} \beta_{2}, \mu_{2} \nu_{2}}(s_{24},t_{2}) \,
\bar{u}(p_{2}, \lambda_{2}) 
i\Gamma^{(\Pom pp)}_{\mu_{2} \nu_{2}}(p_{2},p_{b}) 
u(p_{b}, \lambda_{b}) \,,
\label{amplitude_t}
\end{eqnarray}
\begin{eqnarray} 
{\cal M}^{({\hat{u}})}_{\kappa_{3} \kappa_{4}} 
= 
&& (-i) \bar{u}(p_{1}, \lambda_{1}) 
i\Gamma^{(\Pom pp)}_{\mu_{1} \nu_{1}}(p_{1},p_{a}) 
u(p_{a}, \lambda_{a})\,
i\Delta^{(\Pom)\, \mu_{1} \nu_{1}, \alpha_{1} \beta_{1}}(s_{14},t_{1})  \nonumber \\
&& \times 
i\Gamma^{(\Pom K^{*}K^{*})}_{\kappa_{4} \kappa_{1} \alpha_{1} \beta_{1}}(p_{4},\hat{p}_{u})\,
i\Delta^{(K^{*})\,\kappa_{1} \kappa_{2}}(\hat{p}_{u})\,
i\Gamma^{(\Pom K^{*}K^{*})}_{\kappa_{2} \kappa_{3} \alpha_{2} \beta_{2}}(\hat{p}_{u},-p_{3}) \nonumber \\
&& \times 
i\Delta^{(\Pom)\, \alpha_{2} \beta_{2}, \mu_{2} \nu_{2}}(s_{23},t_{2}) \,
\bar{u}(p_{2}, \lambda_{2}) 
i\Gamma^{(\Pom pp)}_{\mu_{2} \nu_{2}}(p_{2},p_{b}) 
u(p_{b}, \lambda_{b}) \,,
\label{amplitude_u}
\end{eqnarray}
where $\hat{p}_{t} = p_{a} - p_{1} - p_{3}$,
$\hat{p}_{u} = p_{4} - p_{a} + p_{1}$, $s_{ij} = (p_{i} + p_{j})^{2}$.

Our ansatz for the $\Pom K^{*}K^{*}$ vertex follows the one for the $\Pom \rho\rho$ in (3.47) of \cite{Ewerz:2013kda} 
with the replacements 
$a_{\Pom \rho \rho} \to a_{\Pom K^{*}K^{*}}$ and 
$b_{\Pom \rho \rho} \to b_{\Pom K^{*}K^{*}}$;
see also Eqs.~(3.12)--(3.14) of \cite{Lebiedowicz:2019jru}.
With $k', \mu$ and $k,\nu$ the momentum and vector index
of the outgoing and incoming $K^{*}$, respectively,
and $\kappa \lambda$ the tensor-pomeron indices,
the $\Pom K^{*}K^{*}$ vertex 
reads
\begin{eqnarray}
i\Gamma^{(\Pom K^{*}K^{*})}_{\mu \nu \kappa \lambda}(k',k) =
i F_{M}((k'-k)^{2}) \left[
2a_{\Pom K^{*}K^{*}}\,  
\Gamma^{(0)}_{\mu \nu \kappa \lambda}(k',-k)\,
- b_{\Pom K^{*}K^{*}}\,\Gamma^{(2)}_{\mu \nu \kappa \lambda}(k',-k) \right] \,.\quad
\label{vertex_pomKK}
\end{eqnarray}  
Here the coupling parameters $a$ and $b$ have dimensions 
GeV$^{-3}$ and GeV$^{-1}$, respectively.
We take for $F_{M}(t)$ the form given in (\ref{FM_t})
but with $\tilde{\Lambda}_{0}^{2} \to \Lambda_{0}^{2}$,
\begin{eqnarray}
F_{M}(t)=\frac{1}{1-t/\Lambda_{0}^{2}}\,.
\label{FM_t_continuum}
\end{eqnarray}
The amplitudes (\ref{amplitude_t}) and (\ref{amplitude_u})
also contain a form factors for the off-shell dependencies 
of the intermediate $K^{*}$ mesons,
$\hat{F}_{K^{*}}(\hat{p}_{t}^{2})$
and $\hat{F}_{K^{*}}(\hat{p}_{u}^{2})$, respectively.
These form factors are parametrised in the exponential form
\begin{eqnarray} 
\hat{F}_{K^{*}}(\hat{p}^{2})=
\exp\left(\frac{\hat{p}^{2}-m_{K^{*}}^{2}}{\Lambda^{2}_{{\rm off,E}}}\right) \,.
\label{off_shell_form_factors_exp} 
\end{eqnarray}

We assume that only one coupling in (\ref{vertex_pomKK}) contributes,
that is, $a_{\Pom K^{*}K^{*}} \neq 0$ 
or $b_{\Pom K^{*}K^{*}} \neq 0$.
With this assumption, the sign of $a$ or $b$
does not matter as the corresponding coupling occurs twice in the amplitude.
The $\Pom K^{*}K^{*}$ coupling parameters ($a$, $b$)
and the cutoff parameters 
($\Lambda_{0}$, $\Lambda_{\rm off,E}$) 
could be adjusted to experimental data.

For the $K^{*}$-meson propagator 
$\Delta^{(K^{*})}_{\kappa_{1}\kappa_{2}}$ 
using the properties of tensorial functions we can make the replacement
$\Delta_{\kappa_{1}\kappa_{2}}^{(K^{*})}(\hat{p}^{2}) 
\to -g_{\kappa_{1}\kappa_{2}} \Delta^{(K^{*})}_{T}(\hat{p}^{2})$.
We take for $\hat{p}^{2}<0$ the simple expression 
$(\Delta^{(K^{*})}_{T}(\hat{p}^{2}))^{-1} = 
\hat{p}^{2}-m_{K^{*}}^{2}$.

We should take into account the reggeization of 
intermediate $K^{*}$ meson.
In \cite{Harland-Lang:2013dia} it was argued that 
the reggeization should not be applied
when the rapidity distance between two centrally produced mesons, 
$\rm{Y_{diff}} = \rm{Y}_{3} - \rm{Y}_{4}$, tends to zero
\mbox{(i.e. for $|\hat{p}^{2}| \sim s_{34}$)}.
We follow (3.25) of \cite{Lebiedowicz:2019jru} and use
a formula for the $K^{*}$ propagator which interpolates continuously
between the regions of low $\rm{Y_{diff}}$, 
where we use the standard $K^{*}$ propagator, 
and of high $\rm{Y_{diff}}$ where we use the reggeized form:
\begin{eqnarray}
\Delta^{(K^{*})}_{\kappa_{1}\kappa_{2}}(\hat{p})
& \to & \Delta^{(K^{*})}_{\kappa_{1}\kappa_{2}}(\hat{p}) \,F({\rm Y_{diff}})
+ 
\Delta^{(K^{*})}_{\kappa_{1}\kappa_{2}}(\hat{p})\,
\left[ 1 - F({\rm Y_{diff}}) \right]
\left( \exp (i \phi(s_{34}))\, \frac{s_{34}}{s_{\rm thr}} \right)^{\alpha_{K^{*}}(\hat{p}^{2})-1}\,,\nonumber \\
F({\rm Y_{diff}}) &=& \exp\left( -{\rm c_{y}} | {\rm Y_{diff}} | \right)\,,
\nonumber \\
\phi(s_{34}) &=& \frac{\pi}{2}\exp\left(\frac{s_{\rm thr}-s_{34}}{s_{{\rm thr}}}\right)-\frac{\pi}{2}\,,
\label{reggeization}
\end{eqnarray}
where $s_{34} = M_{K^{*0} \bar{K}^{*0}}^{2}$, 
$s_{\rm thr} = 4 m_{K^{*0}}^{2}$, and $\rm{c_{y}}$ is an unknown parameter
which measures how fast one approaches to the Regge regime. 
Here we take ${\rm c_{y}} = 2$. 
This choice is motivated by Fig.~6 of \cite{Lebiedowicz:2019jru}.

We assume for the $K^{*}$ Regge trajectory
a simple linear form 
[see (5.3.1) of \cite{Collins:1977}]
\begin{eqnarray}
\alpha_{K^{*}}(\hat{p}^{2}) = 
\alpha_{K^{*}}(0) + \alpha'_{K^{*}} \,\hat{p}^{2}\,,
\label{Kstar_trajectory_linear}
\end{eqnarray}
with the intercept and slope of the trajectory $\alpha_{K^{*}}(0) = 0.3$ and $\alpha'_{K^{*}} = 0.9$~GeV$^{-2}$, respectively.
We will also show the results using 
a nonlinear Regge trajectory\footnote{For the discussion of nonlinear Regge trajectories 
see Ref.~\cite{Szanyi:2019kkn} and references therein.}
for the $K^{*}$ mesons,
the so-called ``square-root'' trajectory,
parametrised as \cite{Brisudova:1999ut}
\begin{eqnarray}
\alpha_{K^{*}}(\hat{p}^{2}) = 
\alpha_{K^{*}}(0) + 
\gamma \left( \sqrt{T_{K^{*}}} - \sqrt{T_{K^{*}} - \hat{p}^{2}}  \right)\,,
\label{Kstar_trajectory_nonlinear}
\end{eqnarray}
where $\gamma$ governs the slope of the trajectory
and $T_{K^{*}}$ denotes the trajectory termination point.
The parameters are fixed to be 
$\alpha_{K^{*}}(0) = 0.414$,
$\gamma = 3.65$~GeV$^{-1}$, $\sqrt{T_{K^{*}}} = 2.58$~GeV.

\section{Numerical results and discussions}
\label{sec:results}

In this section we wish to present first results for
the $pp \to pp K^{*0}(892) \bar{K}^{*0}(892)$ reaction
and 
for the $pp \to pp K^{+} \pi^{-} K^{-} \pi^{+}$ reaction corresponding to the diagrams in Fig.~\ref{fig:diagrams}.

\subsection{Comparison with the WA102 data}
\label{sec:WA102}

It was noticed by the WA102 Collaboration \cite{Barberis:1998tv} 
that the cross section for the production 
of a $K^{*}(892) \bar{K}^{*}(892)$ system
slowly rises with rising the center-of-mass energy $\sqrt{s}$.
The experimental results, for the same interval 
on the central $K^{*} \bar{K}^{*}$ system
$|x_{F}| \leqslant 0.2$, are
$\sigma_{\rm exp} = 67 \pm 16$~nb at $\sqrt{s} = 12.7$~GeV \cite{Armstrong:1986cm},
$\sigma_{\rm exp} = 70 \pm 14$~nb at $\sqrt{s} = 23.8$~GeV \cite{Armstrong:1989es},
and
$\sigma_{\rm exp} = 85 \pm 10$~nb at $\sqrt{s} = 29.1$~GeV \cite{Barberis:1998tv}.
This suggests that the pomeron-pomeron fusion mechanism is the dominant one
for the $pp \to pp K^{*0} \bar{K}^{*0}$ reaction in the above energy range.
A similar behaviour of the cross section as a function $\sqrt{s}$ 
was observed experimentally also for the $\phi \phi$ production \cite{Barberis:1998bq}.
In the following we neglect, therefore, secondary reggeon exchanges.

\begin{figure}[!ht]
\includegraphics[width=0.46\textwidth]{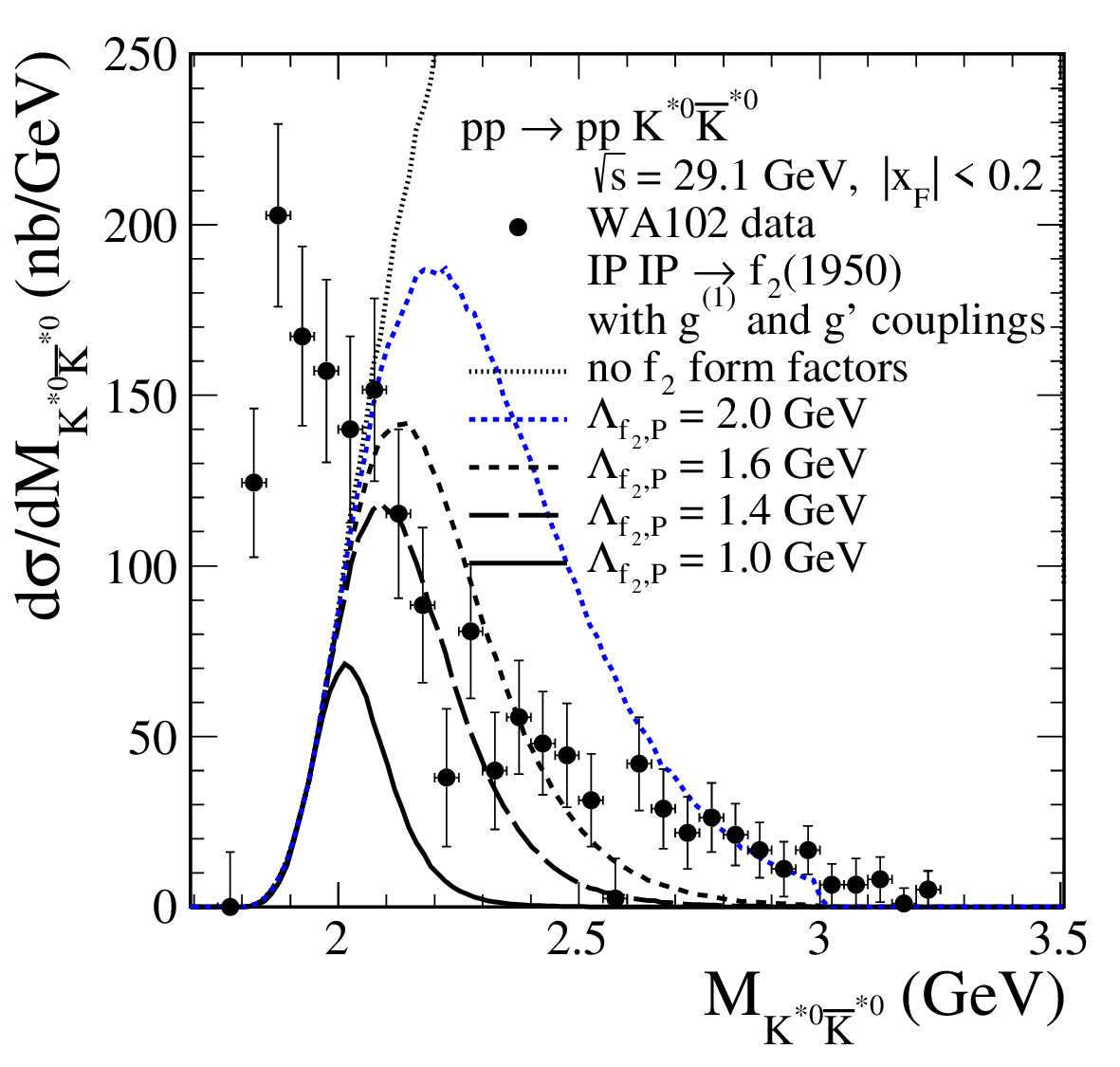}
\includegraphics[width=0.46\textwidth]{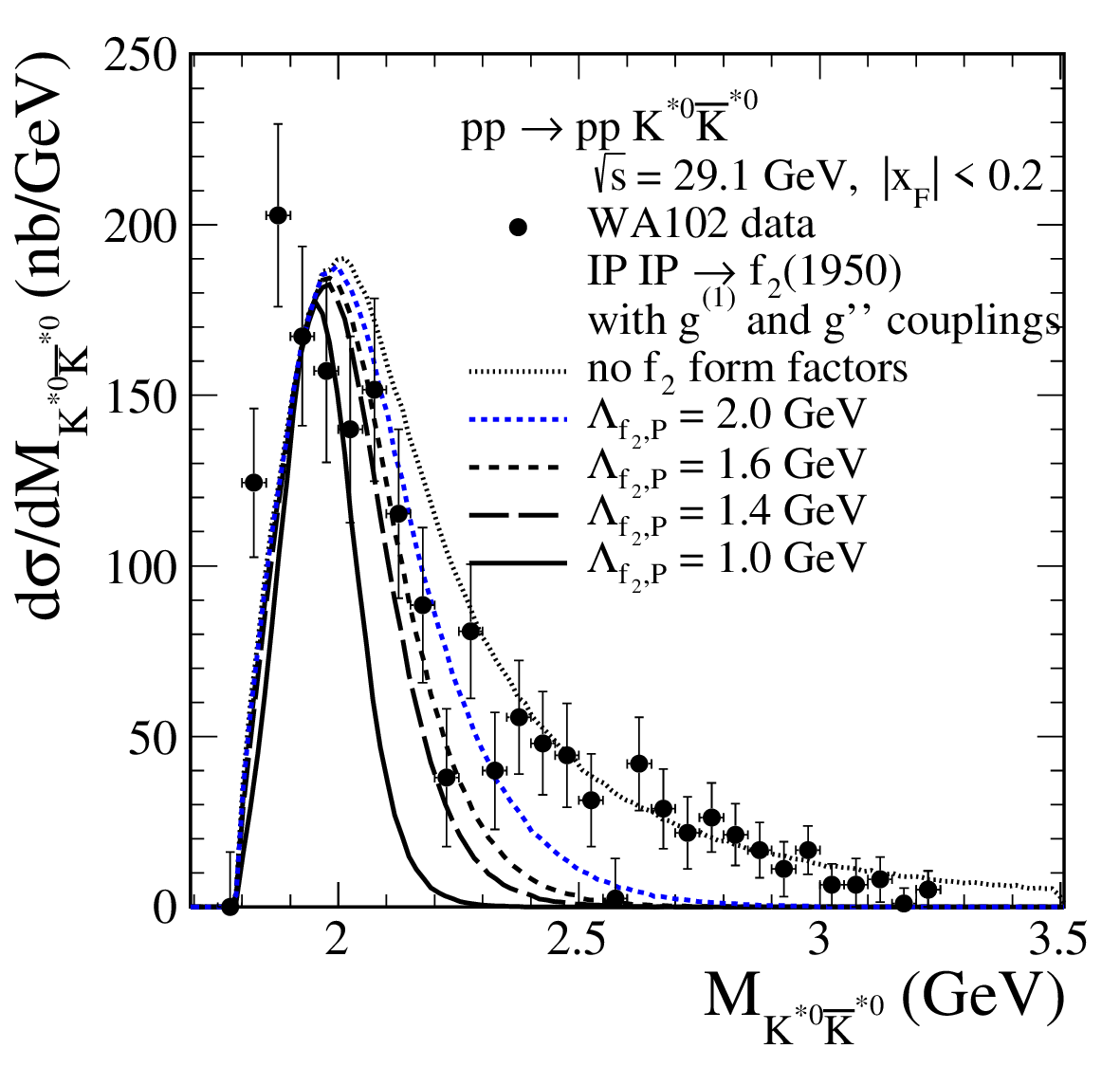}
\caption{\label{fig:1}
The distributions in $K^{*0} \bar{K}^{*0}$ invariant mass 
compared to the WA102 data \cite{Barberis:1998tv} 
for the $\Pom \Pom \to f_{2}(1950)$ contribution.
The calculations were done for $\sqrt{s} = 29.1$~GeV 
and for $|x_{F}| \leqslant 0.2$ of the central $K^{*} \bar{K}^{*}$ system.
The data points have been normalized 
to the total cross section $\sigma_{\rm exp} = 85$~nb.
We show results for the two sets of coupling constants
$|g^{(1)}_{\Pom \Pom f_{2}} g'_{f_{2} K^{*}\bar{K}^{*}}| = 28.0$ (left panel) and 
$|g^{(1)}_{\Pom \Pom f_{2}} g''_{f_{2} K^{*}\bar{K}^{*}}| = 11.0$ (right panel)
and for various cutoff parameters 
$\Lambda_{f_{2},P} = 1.0, 1.4, 1.6$, and 2.0~GeV 
in the form factors (\ref{Fpompommeson_ff_P}) 
and (\ref{Fpompommeson_ff_tensor}) describing
the off-shellnes of the $f_{2}$ meson.
We have taken here $\tilde{\Lambda}_{0}^{2} = 0.5\;{\rm GeV}^{2}$ (\ref{FM_t}).
In addition,
we show also a naive results that corresponds 
to the calculations without these form factors.
The absorption effects are included.}
\end{figure}

In Fig.~\ref{fig:1} we show the invariant mass distributions
for the $\Pom \Pom \to f_{2}(1950)$ mechanism together with
the WA102 experimental data from Fig.~2 of \cite{Barberis:1998tv}.
The data points have been normalised to the mean value of the
total cross section
$\sigma_{\rm exp} = 85 \pm 10$~nb from \cite{Barberis:1998tv}.
For the purpose of orientation,
we have assumed, 
that in the $\Pom \Pom f_{2}(1950)$ vertex (\ref{vertex_pompomT})
only $g^{(1)}$ coupling constant is unequal to zero.
We have checked that for the distributions studied here
the choice of $\Pom \Pom f_{2}$ coupling is not important.
This is similar to what was found in \cite{Lebiedowicz:2019jru} 
for the reaction $pp \to pp (\Pom \Pom \to f_{2}(2340) \to \phi \phi)$.
In the calculation we take only one $\Pom \Pom f_{2}$ coupling  
[$g^{(1)}_{\Pom \Pom f_{2}}$ from (\ref{vertex_pompomT})]
and only one $f_{2} K^{*}\bar{K}^{*}$ coupling
[$g'_{f_{2} K^{*}\bar{K}^{*}}$ or $g''_{f_{2} K^{*}\bar{K}^{*}}$
from (\ref{vertex_f2KKbar})].
The results shown in the left panel 
correspond to the product of the couplings 
$|g^{(1)} \times g'| = 28.0$, 
while the results in the right panel are for
$|g^{(1)} \times g''| = 11.0$.
We note that only the absolute value of both products
for fixed the cutoff parameter 
of the $f_{2}$-meson off-shell form factor
(\ref{Fpompommeson_ff_tensor}) can be determined.
We have checked that in both cases
the results for the product of the form factors
$F^{(\Pom \Pom f_{2})}(p_{34}^{2}) \times
F^{(f_{2} K^{*}\bar{K}^{*}}(p_{34}^{2})$ 
assuming the same type of form factors,
(\ref{Fpompommeson_ff_E}) or (\ref{Fpompommeson_ff_P}),
are similar. 
In the following we choose in the calculation
only the power form (\ref{Fpompommeson_ff_P})
with the cutoff parameter $\Lambda_{f_{2},P}$ (\ref{vertex_f2KKbar}).
It is clearly seen from the left panel that
the result without these form factors, i.e.,
for $p_{34}^{2} = m_{f_{2}}^{2}$, is well above
the WA102 experimental data for $M_{K^{*0} \bar{K}^{*0}} > 2.1$~GeV.
The results are very sensitive to the 
choice of the cutoff parameters.
It can be observed that as $\Lambda_{f_{2},P}$ decreases
then mainly the right flank of the resonance is reduced
and thus it becomes narrower.
For $\Lambda_{f_{2},P} = 1.6 - 2.0$~GeV
and $|g^{(1)} \times g''| = 11.0$
(see the short-dashed lines in the right panel)
an agreement with the WA102 data 
in the invariant mass range
$M_{K^{*0} \bar{K}^{*0}} \in (1.9, 2.2)$~GeV
can be obtained.

\begin{figure}[!ht]
\includegraphics[width=0.46\textwidth]{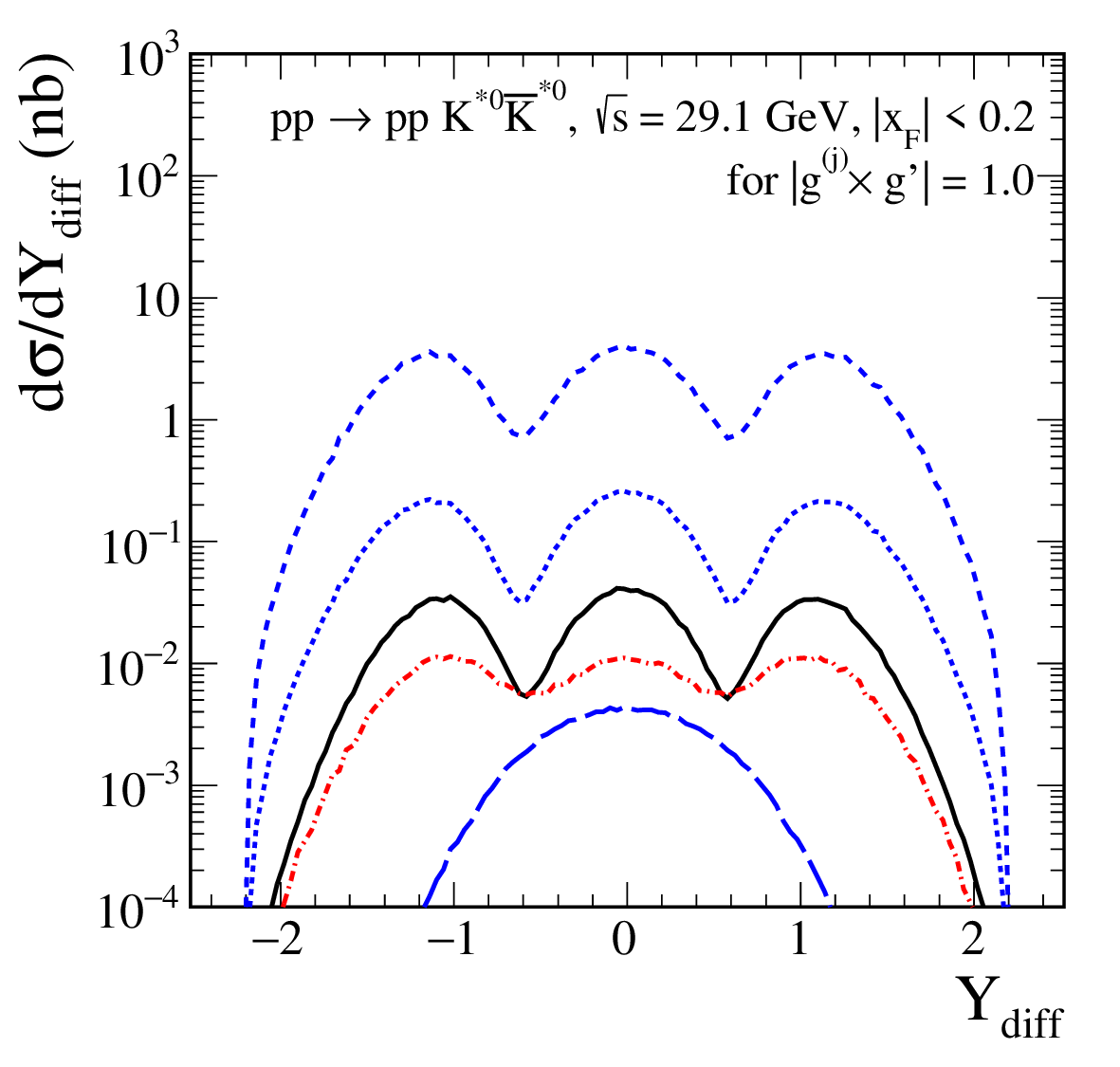}
\includegraphics[width=0.46\textwidth]{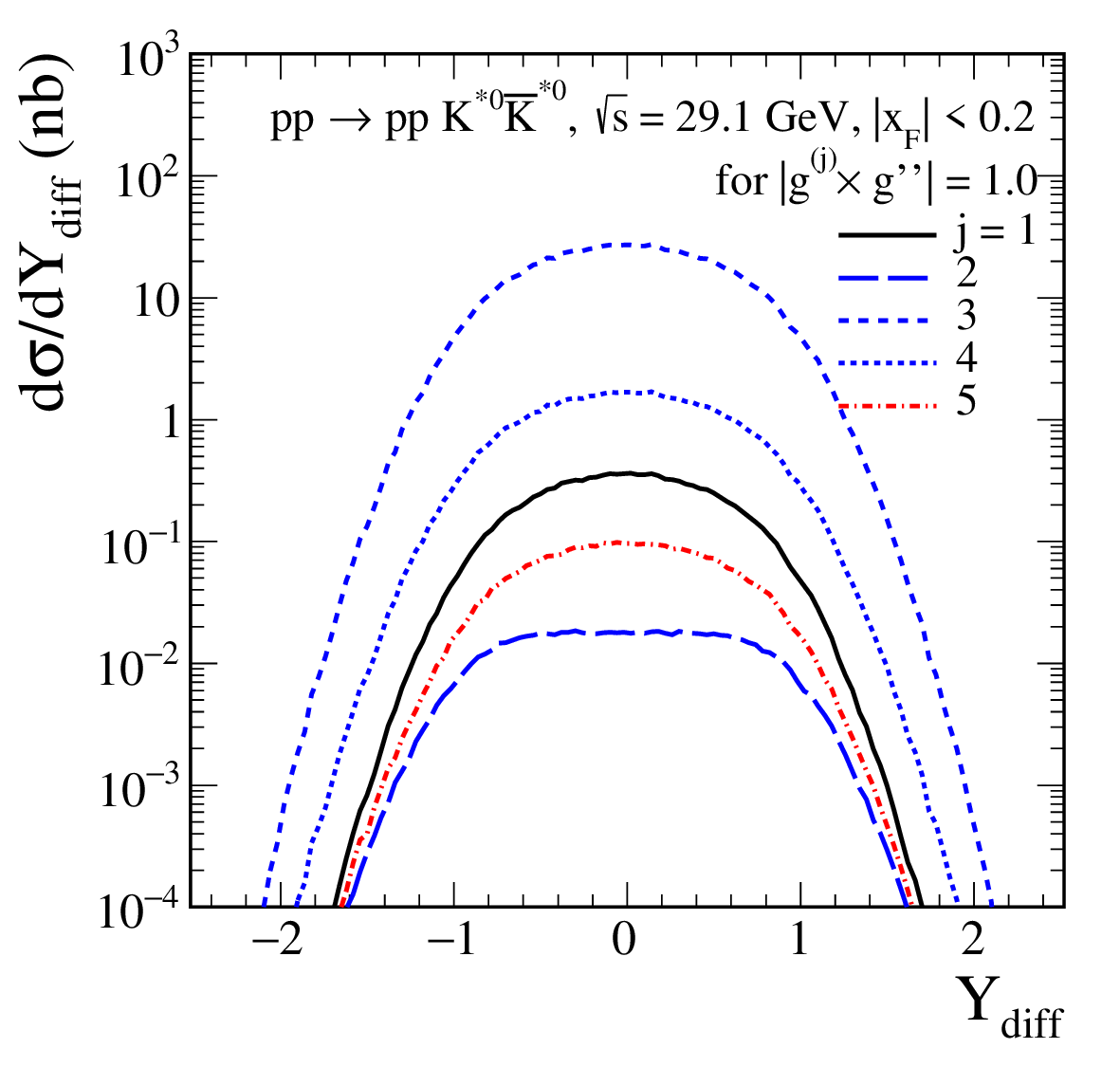}
\caption{\label{fig:1a}
The distributions in $\rm{Y_{diff}}$ 
for the process $\Pom \Pom \to f_{2}(1950) \to K^{*0} \bar{K}^{*0}$.
The calculations were done for $\sqrt{s} = 29.1$~GeV 
and for $|x_{F}| \leqslant 0.2$ of the central $K^{*} \bar{K}^{*}$ system.
We show the individual contributions of the different $\Pom \Pom f_{2}$ 
couplings (\ref{vertex_pompomT}) with index $j$.
We have taken here $\tilde{\Lambda}_{0}^{2} = 0.5\;{\rm GeV}^{2}$
and $\Lambda_{f_{2},P} = 1.6$~GeV.
The results in the left panel have been obtained with coupling constants
$|g^{(j)}_{\Pom \Pom f_{2}} g'_{f_{2} K^{*}\bar{K}^{*}}| = 1.0$,
while the results in the right panel with
$|g^{(j)}_{\Pom \Pom f_{2}} g''_{f_{2} K^{*}\bar{K}^{*}}| = 1.0$.
The absorption effects are included.}
\end{figure}

\begin{figure}[!ht]
\includegraphics[width=0.46\textwidth]{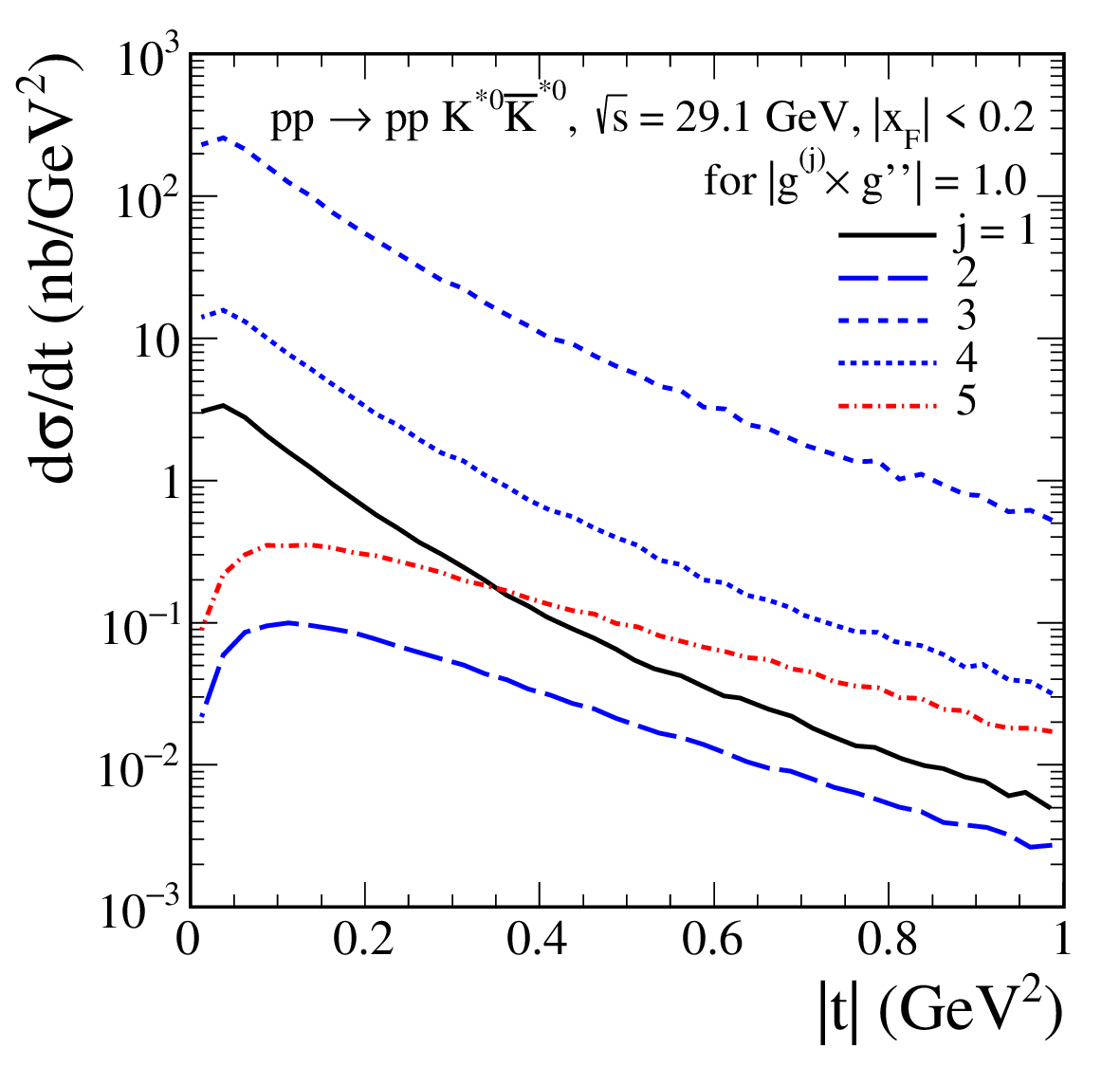}
\includegraphics[width=0.46\textwidth]{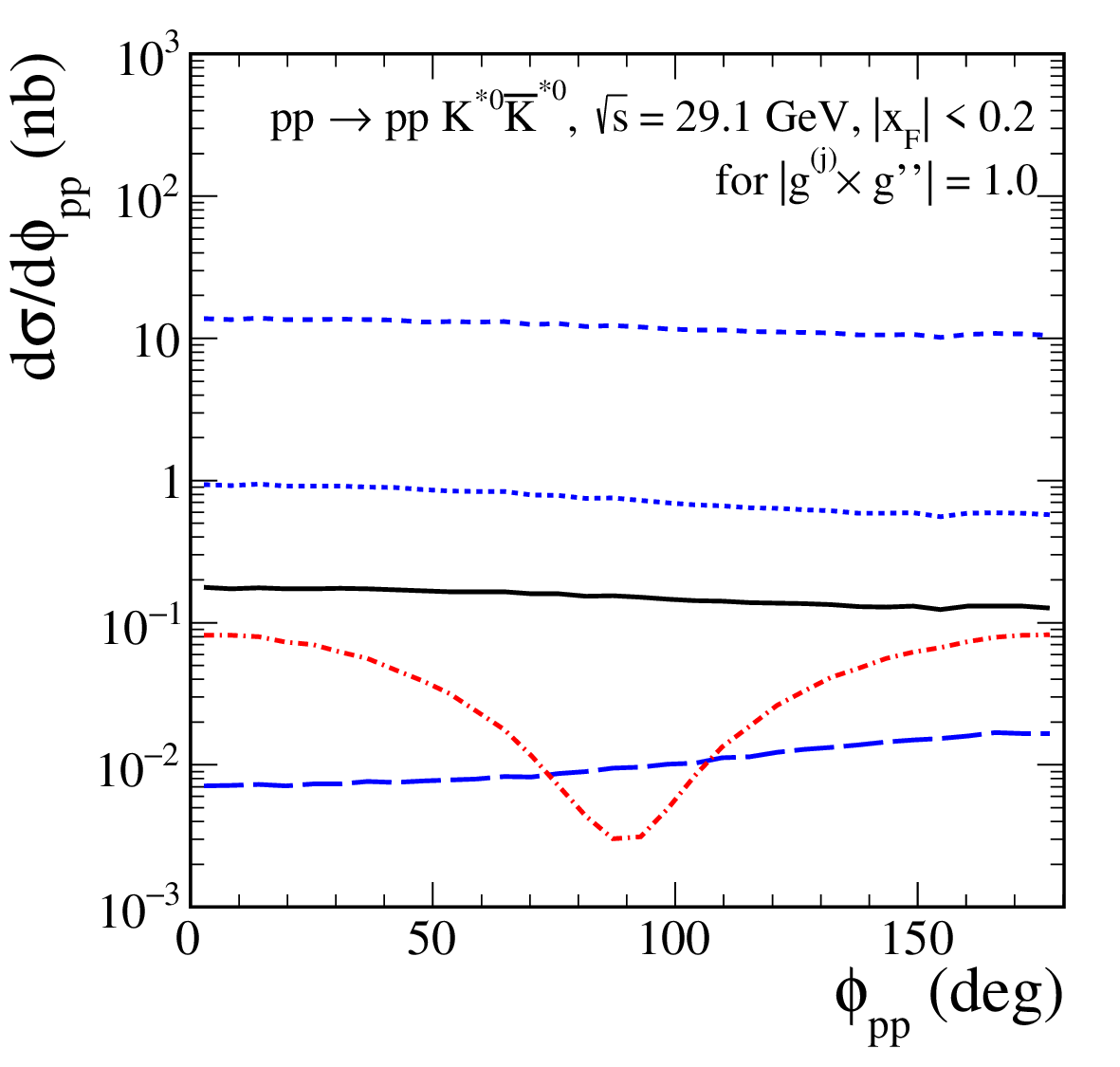}
\caption{\label{fig:1b}.
The $|t|$ (left panel) and $\phi_{pp}$ (right panel) distributions 
for the $pp \to pp (\Pom \Pom \to f_{2}(1950) \to K^{*0} \bar{K}^{*0})$ reaction for the WA102 kinematics.
The meaning of the lines is the same as 
in the right panel of Fig.~\ref{fig:1a}.
The calculation was done for
$|g^{(j)}_{\Pom \Pom f_{2}} g''_{f_{2} K^{*}\bar{K}^{*}}| = 1.0$.
The absorption effects are included.}
\end{figure}

In Figs.~\ref{fig:1a} and \ref{fig:1b} we show different differential
observables in $\rm{Y_{diff}}$, 
the rapidity difference between the two $K^{*0}$ mesons,
in $|t|$, 
the transferred four-momentum squared 
from one of the proton vertices ($t = t_{1}$ or $t_{2}$),
and in $\phi_{pp}$, the azimuthal angle between 
the transverse momentum vectors $\bpta$ and $\bptb$ of the outgoing protons.
We present the results obtained separately for different couplings
taking into account the absorptive corrections.
In the left panel of Fig.~\ref{fig:1b} we show results for
the individual $j$ coupling terms 
$g^{(j)}_{\Pom \Pom f_{2}} \times g'_{f_{2} K^{*}\bar{K}^{*}}$
(only for five terms),
while in the right panel for $g^{(j)}_{\Pom \Pom f_{2}} \times g''_{f_{2} K^{*}\bar{K}^{*}}$.
For illustration, the results have been obtained
with coupling constants
$|g^{(j)} \times g'| = 1.0$ (left panel) and $|g^{(j)} \times g''| = 1.0$ (right panel).
The shape of the $\rm{Y_{diff}}$ distribution 
depends on the choice 
of the $f_{2}(1950) K^{*}\bar{K}^{*}$ coupling.
It can be expected that this variable 
will be very helpful in determining 
the $f_{2} K^{*} \bar{K}^{*}$ coupling using data
from LHC measurements, in particular, 
if they cover a wider range of rapidities;
see the discussion 
in Sec.~IV~B of Ref.~\cite{Lebiedowicz:2019jru}.
The shapes of the distributions in $\rm{Y_{diff}}$ 
within each group are similar except of $j = 2$ term.
In Fig.~\ref{fig:1b} we show the results only for the second group
with the $g''$ coupling.
We have checked that with the $g'$ coupling
the shapes of the distributions for these observables are very similar.
Compared to the WA102 data from \cite{Barberis:1998tv}
that will be presented later (Fig.~\ref{fig:3}), 
it can be concluded that the terms $j = 2$ and 5
can be excluded.
We find that the three cases, $j = 1, 3$ and 4,
give similar characteristics for the WA102 data.
In the following considerations, 
for simplicity,
we assume only one set of couplings, 
namely, $j =1$ $g^{(1)}_{\Pom \Pom f_{2}}$ 
and $g''_{f_{2} K^{*}\bar{K}^{*}}$.

\begin{figure}[!ht]
\includegraphics[width=0.46\textwidth]{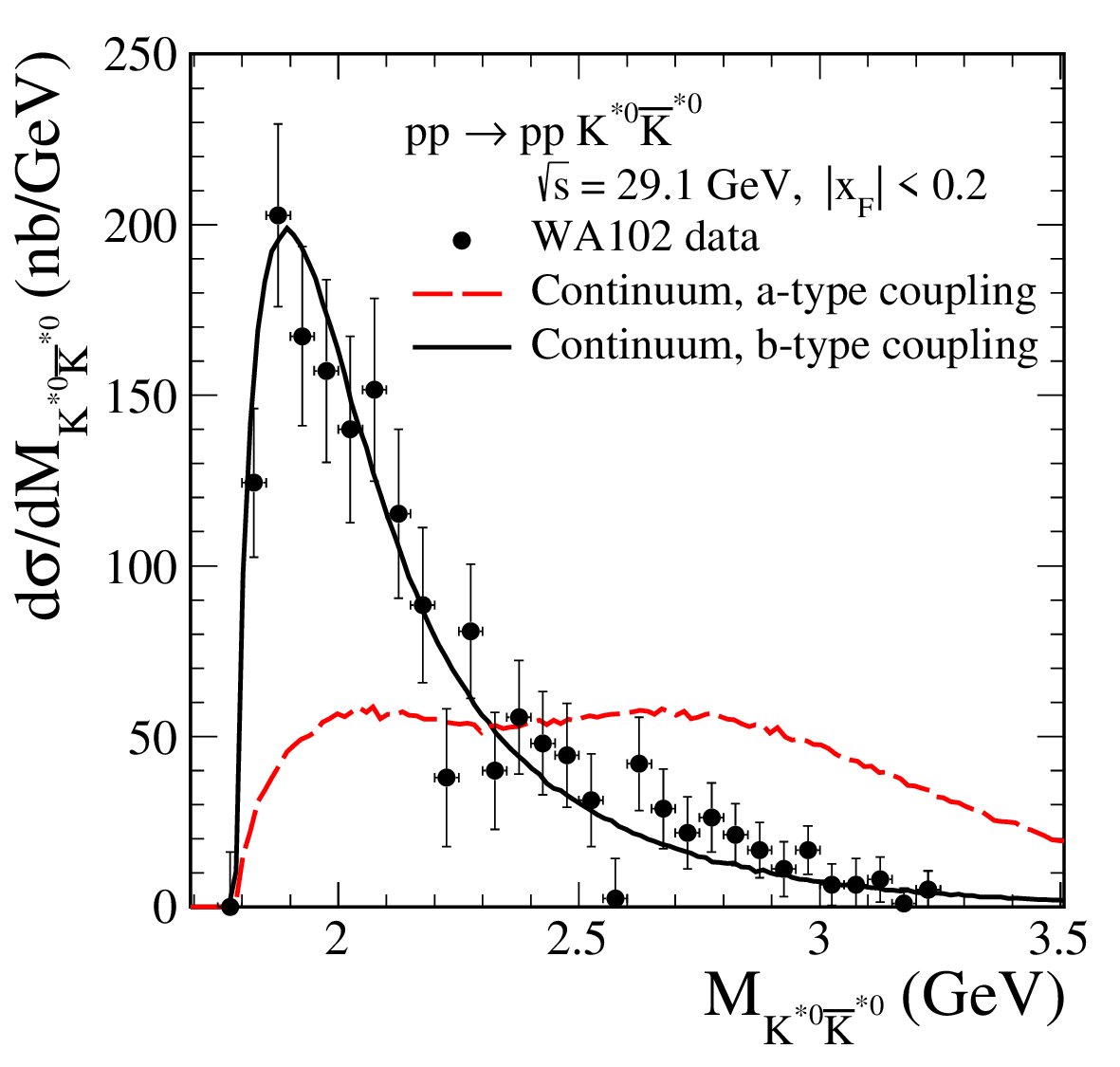}
\caption{\label{fig:2}
The same as in Fig.~\ref{fig:1} but here we show 
the theoretical results for the continuum mechanism.
We show results for the two type of the $\Pom K^{*} K^{*}$ coupling 
considered separately,
$a_{\Pom K^{*}K^{*}}$ and $b_{\Pom K^{*}K^{*}}$, 
occurring in (\ref{vertex_pomKK}).
The results are normalized to the same value $\sigma = 85$~nb.
The red dashed line corresponds to 
$|a| = 1.83$~GeV$^{-3}$ and $b = 0$,
while the black solid line corresponds to $a = 0$ and
$|b| = 4.37$~GeV$^{-1}$.
The absorption effects are included.}
\end{figure}

\begin{figure}[!ht]
\includegraphics[width=0.46\textwidth]{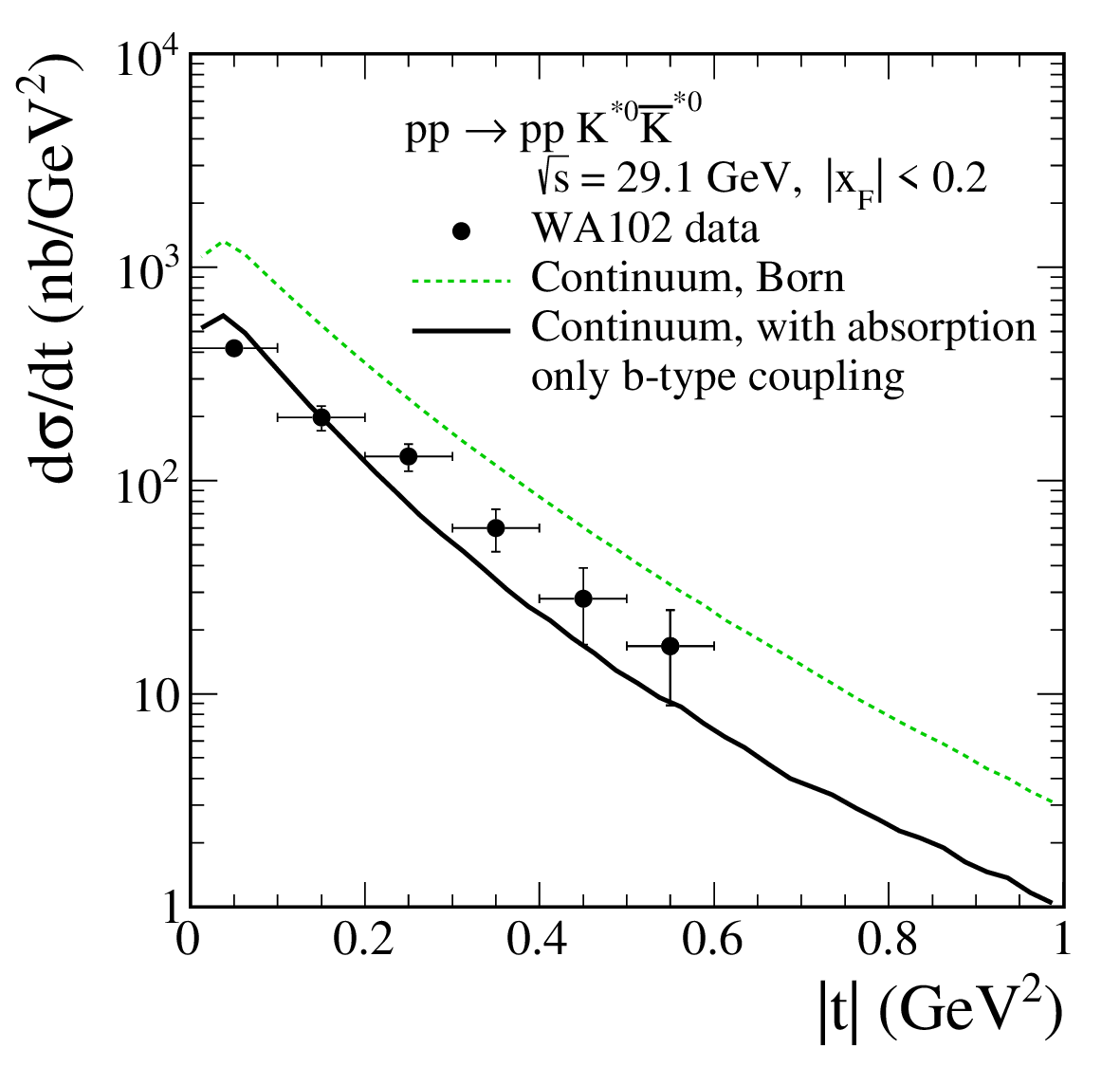}
\includegraphics[width=0.46\textwidth]{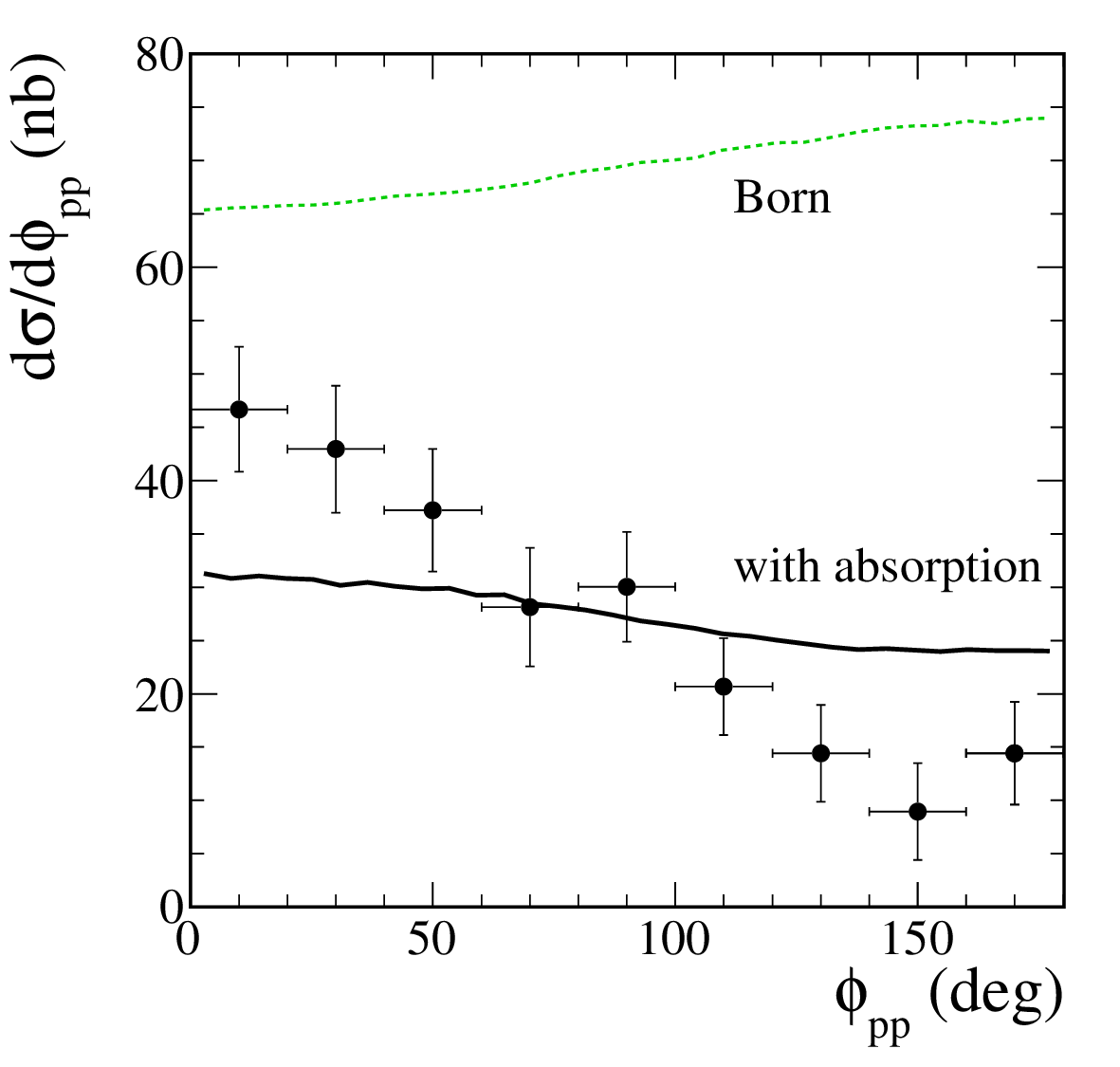}
\caption{\label{fig:3}
The $|t|$ (left panel) and $\phi_{pp}$ (right panel) distributions 
for the $pp \to pp K^{*0} \bar{K}^{*0}$ reaction 
at $\sqrt{s} = 29.1$~GeV and $|x_{F,K^{*} \bar{K}^{*}}| \leqslant 0.2$.
The data points from \cite{Barberis:1998tv}
have been normalised to the total cross section 
$\sigma_{\rm exp} = 85$~nb.
We show results for the continuum contribution
obtained with the $b$-type coupling only
in the Born approximation and with absorption.}
\end{figure}

Now we turn to the diffractive continuum mechanism.
In Fig.~\ref{fig:2} we show the results
for the continuum process via the $K^{*0}$-meson exchange
including the reggeization effect given in (\ref{reggeization}), (\ref{Kstar_trajectory_linear}).
In our calculation we take $\Lambda_{\rm off,E} = 1.6$~GeV
in (\ref{off_shell_form_factors_exp}).
We compare our results assuming only one type of
the $\Pom K^{*} K^{*}$ coupling, $a$ or $b$, 
to the WA102 experimental data
for the $M_{K^{*0} \bar{K}^{*0}}$ distribution.
Our model calculation with only the $b$-type coupling
($a = 0$ and $|b| = 4.37$~GeV$^{-1}$)
describes the experimental data reasonably well,
although, because of large experimental error bars,
a small contribution from the $a$-type coupling 
cannot be ruled out.The option $|a| = 1.83$~GeV$^{-3}$ 
and $b = 0$ (see the dashed line)
is clearly ruled out by the WA102 data.
We cannot also completely rule out 
some contribution of the $f_{2}(1950)$ resonance.
We wish to point out that the interference effects possible 
between these terms may also play an important role;
see \cite{Lebiedowicz:2019jru}.
This requires further analysis and will only be meaningful once
experiments with better statistics become available.
Hopefully this will be the case at the LHC.

In Fig.~\ref{fig:3} 
we show the results for the $|t|$ and $\phi_{pp}$ distributions
together with the experimental data from Fig.~3 of \cite{Barberis:1998tv}.
The data points have been normalised to the mean value of the total cross section ($\sigma_{{\rm exp}} = 85 \pm 10$~nb)
from \cite{Barberis:1998tv}.
We present results only for the continuum
$K^{*0}$-exchange contribution without (the top lines)
and with (the bottom lines) the absorption
effects included in the calculations.
We have checked that the $f_{2}(1950)$-exchange contribution
(with $g^{(1)}$ and $g'$ or $g''$ couplings)
has a very similar shape of these distributions.
The absorption effects lead to a large reduction of the cross section.
We can see a large damping of the cross section 
in the region of $\phi_{pp} \sim \pi$.
The ratio of full (including absorption) and Born cross sections
$\langle S^{2} \rangle$, the gap survival factor, 
for the WA102 kinematics
($\sqrt{s} = 29.1$~GeV and $|x_{F,K^{*} \bar{K}^{*}}| \leqslant 0.2$)
is $\langle S^{2} \rangle \cong 0.4$ for the continuum contribution
and $0.38$ for the $f_{2}$ contribution.

In \cite{Barberis:1998tv} also the ${\rm dP_{t}}$ dependence
for the $K^{*} \bar{K}^{*}$ system was presented.
Here, ${\rm dP_{t}}$ (the so-called ``glueball-filter variable'' 
\cite{Close:1997pj,Barberis:1996iq}) is defined as 
\begin{eqnarray}
\bdPt = \bqta - \bqtb = \bptb - \bpta \,, 
\quad {\rm dP_{t}} = |\bdPt|\,.
\label{dPt_variable}
\end{eqnarray}

In Table~\ref{tab:ratio_dPt} we show the WA102 experimental values 
for the fraction of $K^{*} \bar{K}^{*}$ production
in three ${\rm dP_{t}}$ intervals and for
the ratio of production at small 
${\rm dP_{t}}$ to large ${\rm dP_{t}}$
and our corresponding results for 
the $f_{2}(1950)$ meson and continuum contributions.
The calculations have been done with the absorption effects included.
From the comparison to the WA102 results we see 
that smaller values of the cutoff parameter,
$\tilde{\Lambda}_{0}^{2} = 0.5$~GeV$^{2}$ in (\ref{FM_t}) and
$\Lambda_{0}^{2} = 0.5$~GeV$^{2}$ in (\ref{FM_t_continuum}),
are preferred.
We can conclude that both the continuum contribution
with the $b$-type $\Pom K^{*}K^{*}$ coupling
and the $f_{2}$ contribution with the $g^{(1)}$ and $g''$ couplings
have similar characteristics as the WA102 data. 
\begin{table}[!ht]
\caption{Results of $K^{*} \bar{K}^{*}$ production 
as a function of ${\rm dP_{t}}$
(\ref{dPt_variable}), in three ${\rm dP_{t}}$ intervals,
expressed as a percentage of the total contribution 
at the WA102 collision energy $\sqrt{s}=29.1$~GeV 
and for $|x_{F,K^{*} \bar{K}^{*}}| \leqslant 0.2$.
In the last column the ratios of 
$\sigma ({\rm dP_{t}} \leqslant \,0.2~\mathrm{GeV})/
\sigma ({\rm dP_{t}} \geqslant \,0.5~\mathrm{GeV})$ are given.
The experimental numbers are from \cite{Barberis:1998tv}. 
The theoretical numbers correspond 
to the $f_{2}(1950)$ production mechanism
with the $g^{(1)} \times g'$ and $g^{(1)} \times g''$ couplings,
$\Lambda_{f_{2},P} = 1.6$~GeV in (\ref{Fpompommeson_ff_P}), 
and $\tilde{\Lambda}_{0}^{2} = 0.5, 1.0\;{\rm GeV}^{2}$ in (\ref{FM_t}).
For the continuum mechanism, we show the results
only with the $b$-type $\Pom K^{*}K^{*}$ coupling, 
$\Lambda_{\rm off,E} = 1.6$~GeV in (\ref{off_shell_form_factors_exp}),
and $\Lambda_{0}^{2} = 0.5, 1.0\;{\rm GeV}^{2}$ in (\ref{FM_t_continuum}).
The absorption effects have been included in our analysis.}
\label{tab:ratio_dPt}
\begin{tabular}{|c|c|c|c|c|}
\hline
 & 
${\rm dP_{t}} \leqslant 0.2$~GeV & 
$0.2 \leqslant {\rm dP_{t}} \leqslant 0.5$~GeV & 
${\rm dP_{t}} \geqslant 0.5$~GeV & Ratio\\
\hline
Experiment \cite{Barberis:1998tv}& 
$23 \pm 3$ & $54 \pm 3$ & $23 \pm 2$ & $1.00 \pm 0.16$\\
\hline
$f_{2}(1950)$,
$g^{(1)}$ and $g'$ couplings& & & &\\
\hline
$\tilde{\Lambda}_{0}^{2} = 0.5\;{\rm GeV}^{2}$& 
20.9 & 56.3 & 22.8 & 0.92 \\
$\tilde{\Lambda}_{0}^{2} = 1.0\;{\rm GeV}^{2}$& 
17.8 & 53.3 & 28.9 & 0.62 \\
\hline
$f_{2}(1950)$,
$g^{(1)}$ and $g''$ couplings& & & &\\
\hline
$\tilde{\Lambda}_{0}^{2} = 0.5\;{\rm GeV}^{2}$& 
20.8 & 56.2 & 23.0 & 0.91 \\
$\tilde{\Lambda}_{0}^{2} = 1.0\;{\rm GeV}^{2}$& 
17.7 & 53.1 & 29.2 & 0.61 \\
\hline
Continuum, $b$-type coupling& & & &\\
\hline
$\Lambda_{0}^{2} = 0.5\;{\rm GeV}^{2}$& 
20.3 & 55.2 & 24.5 & 0.83 \\
$\Lambda_{0}^{2} = 1.0\;{\rm GeV}^{2}$& 
17.1 & 51.8 & 31.0 & 0.55 \\
\hline
\end{tabular}
\end{table}

By comparing the theoretical results and the differential cross sections
obtained by the WA102 Collaboration we fixed the parameters of our model.
With them we will provide our predictions for the LHC.
For the continuum term we take
$|b_{\Pom K^{*}K^{*}}| = 4.37$~GeV$^{-1}$,
$a_{\Pom K^{*}K^{*}} = 0$, 
$\Lambda_{\rm off,E} = 1.6$~GeV, 
$\Lambda_{0}^{2} = 0.5$~GeV$^{2}$,
and for the $f_{2}(1950)$ term we take
$|g^{(1)}_{\Pom \Pom f_{2}} g''_{f_{2} K^{*}\bar{K}^{*}}| = 11.0$,
$\Lambda_{f_{2},P} = 1.6, 2.0$~GeV, 
$\tilde{\Lambda}_{0}^{2} = 0.5$~GeV$^{2}$.

In the future the model parameters 
(coupling constants, form-factor cutoff parameters)
could be verified or, if necessary, adjusted 
by a comparison to precise experimental data 
from the LHC experiments.

\subsection{Predictions for the LHC experiments}
\label{sec:LHC}

Here we shall give our predictions for the reaction
$pp \to pp K^{+}\pi^{-}K^{-}\pi^{+}$
represented by the diagrams in Fig.~\ref{fig:diagrams}.
The results were obtained in the calculations
with the tensor-pomeron exchanges including
the absorptive corrections within the one-channel-eikonal approach.

\begin{figure}[!ht]
\includegraphics[width=0.48\textwidth]{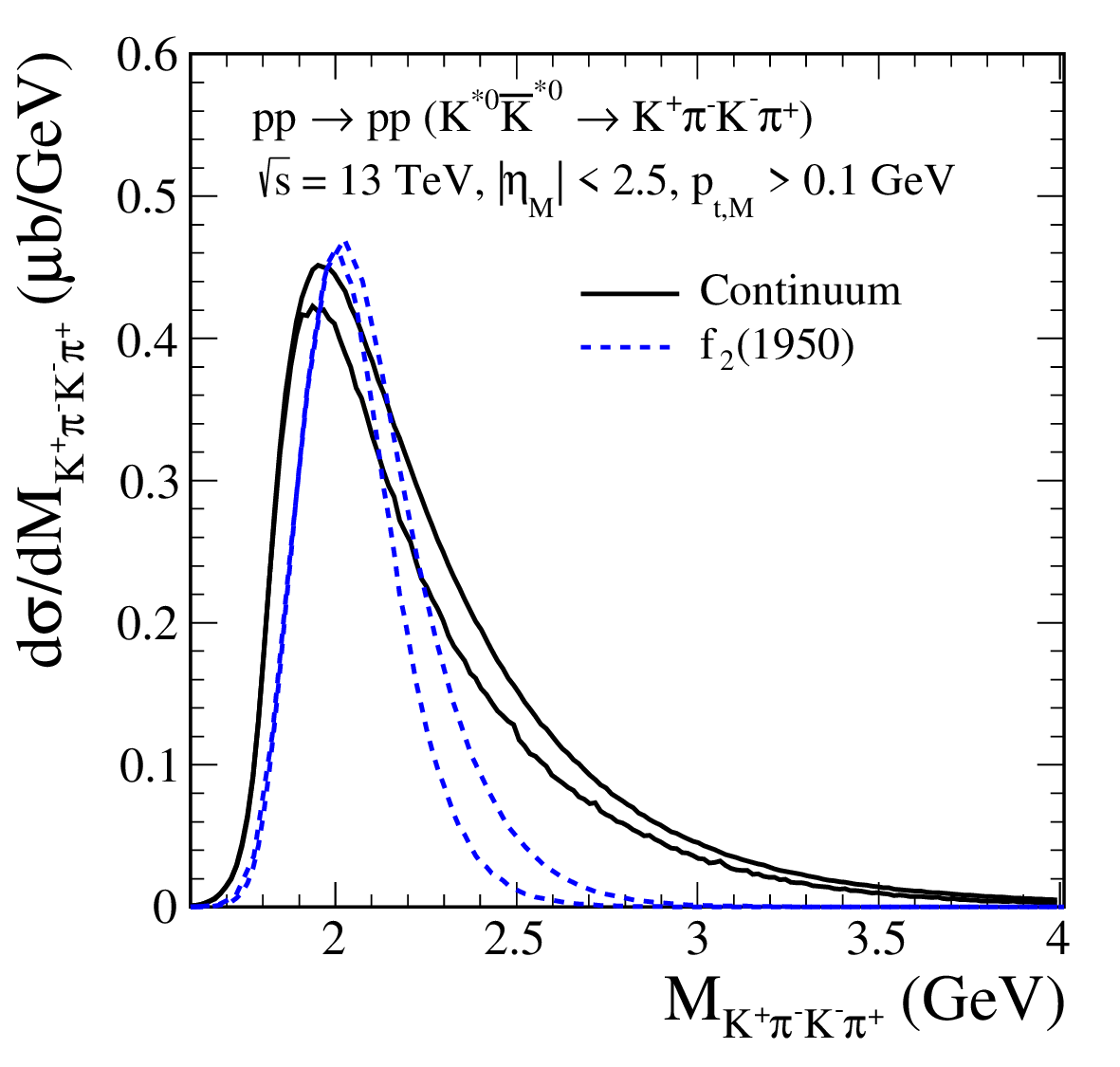}
\includegraphics[width=0.48\textwidth]{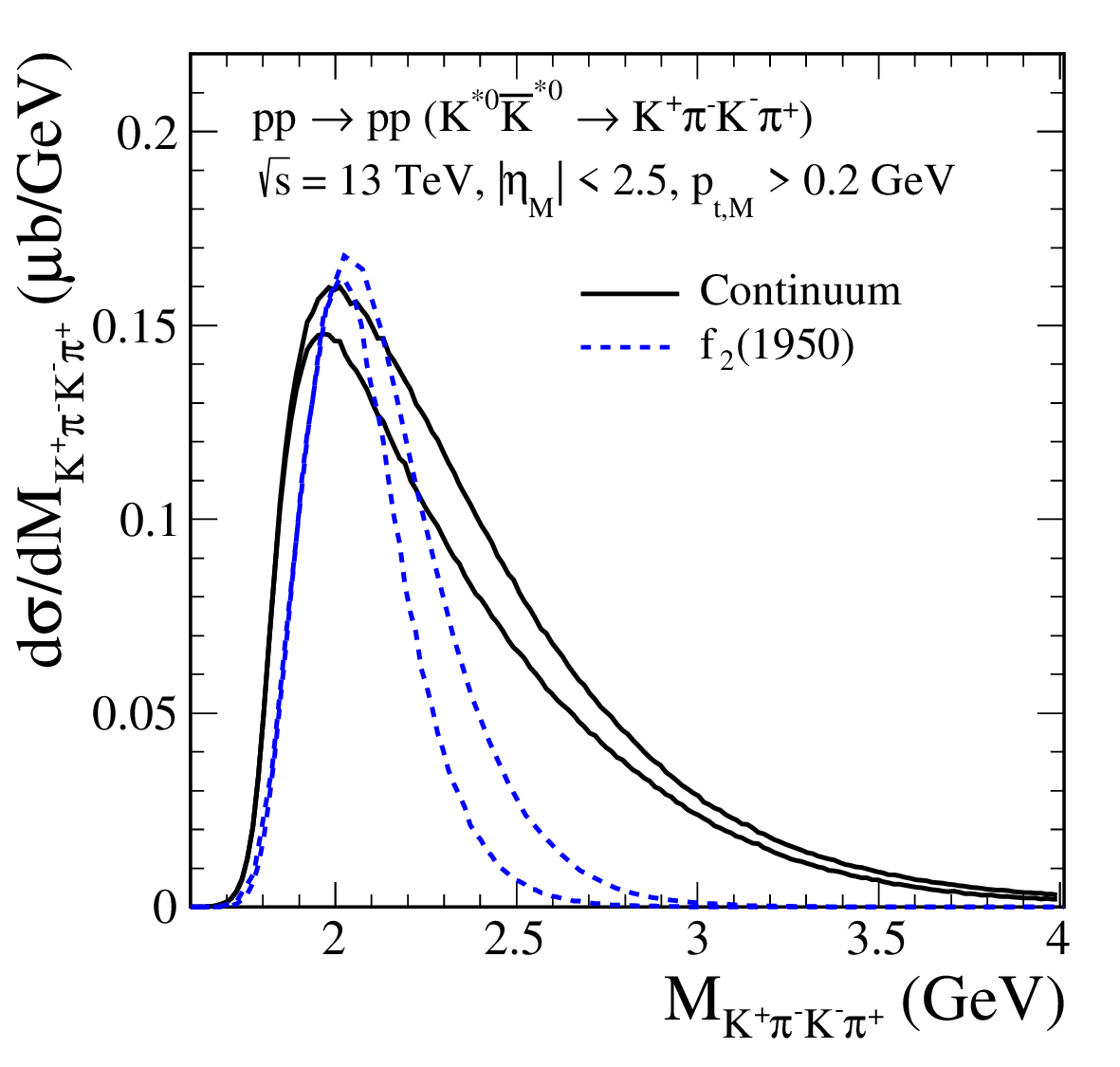}
\includegraphics[width=0.48\textwidth]{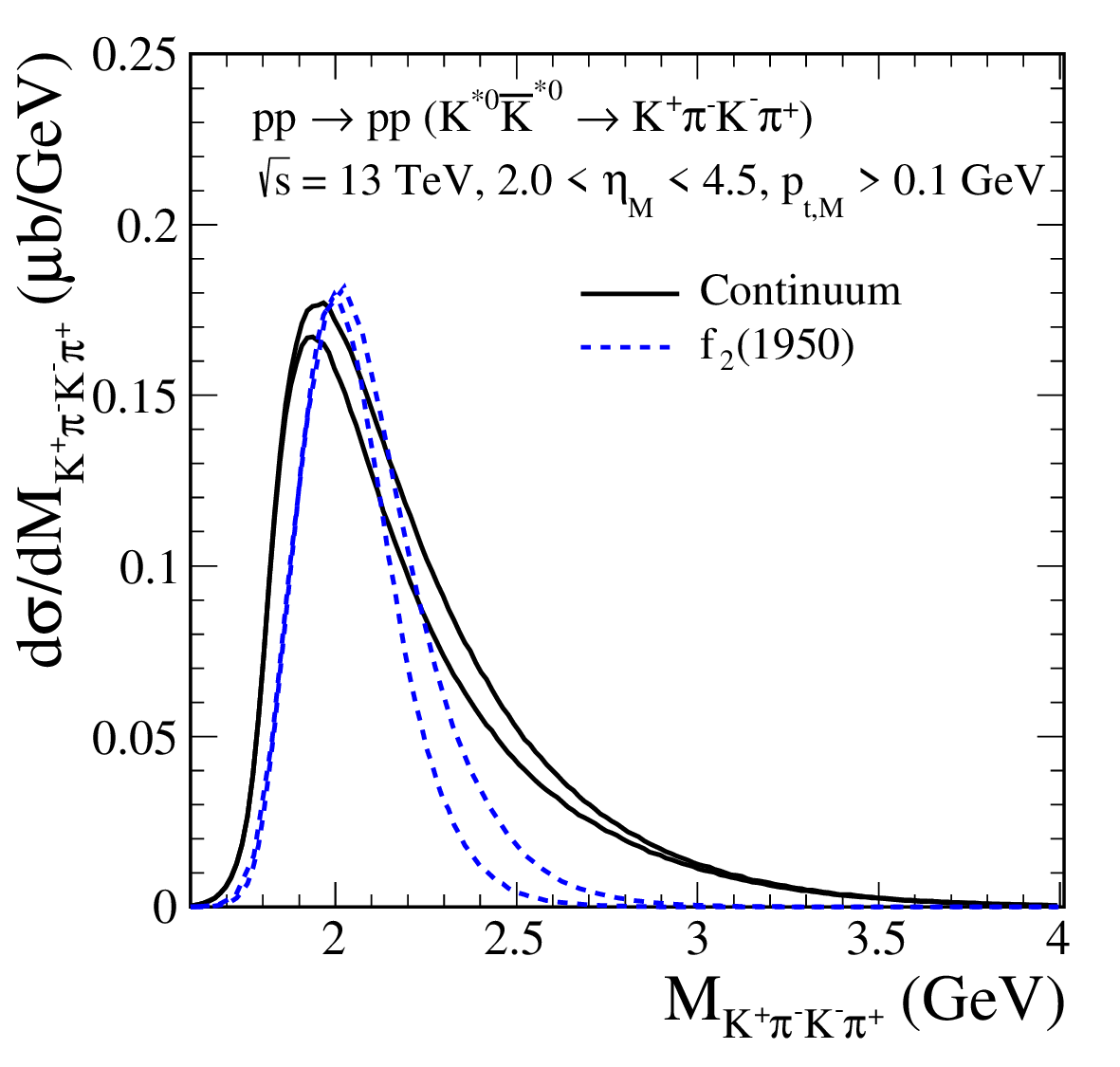}
\includegraphics[width=0.48\textwidth]{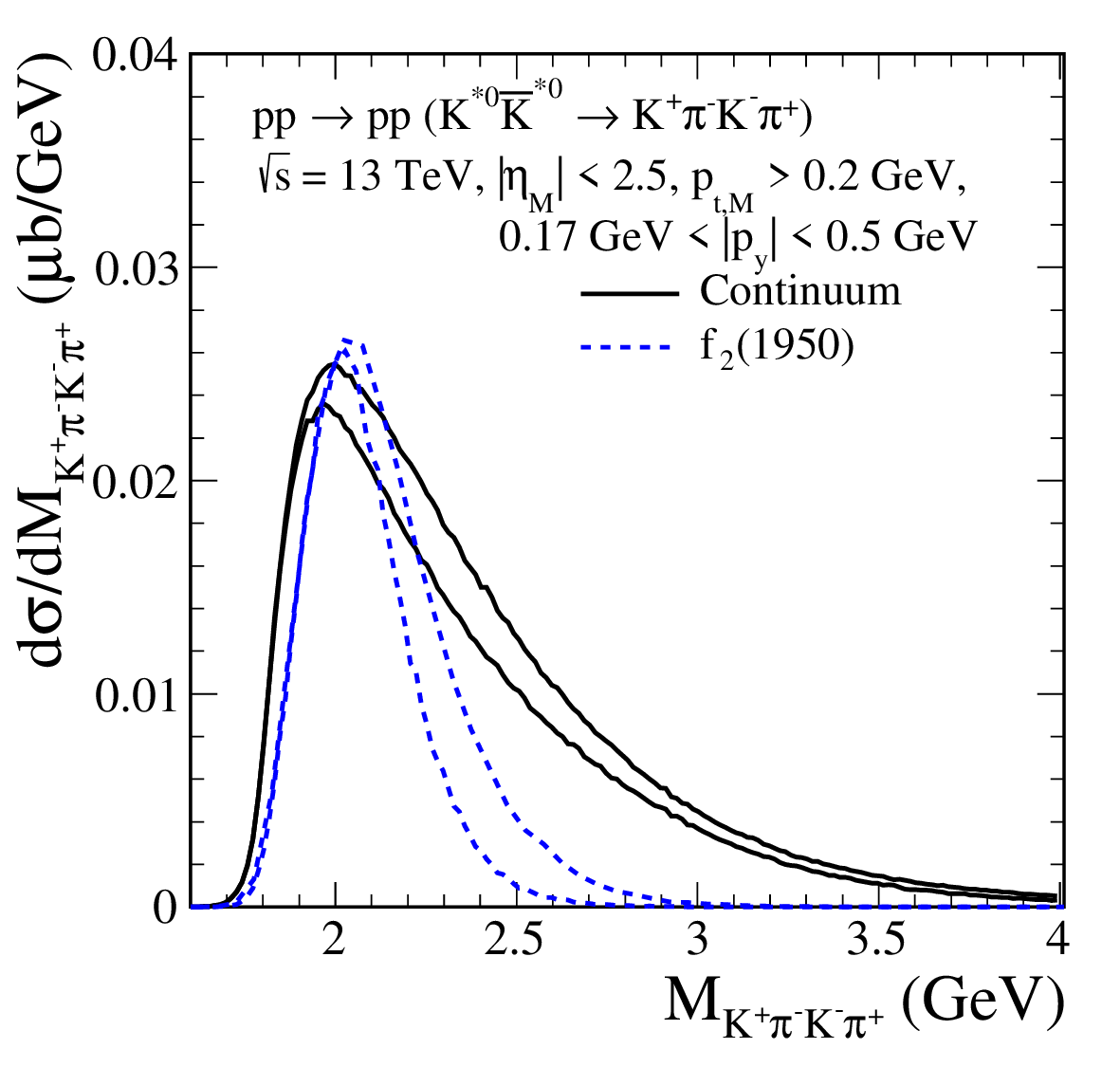}
\caption{\label{fig:4}
Invariant mass distributions for the central $K^{+}\pi^{-}K^{-}\pi^{+}$ system
via the $K^{*0} \bar{K}^{*0}$ states
calculated for $\sqrt{s} = 13$~TeV 
with the kinematical cuts specified in the figure legends.
The results for the two mechanisms are presented.
For the $f_{2}(1950)$ term we show the results for 
$\Lambda_{f_{2},P} = 1.6$~GeV (the lower dashed lines) 
and 2.0~GeV (the upper dashed lines).
For the continuum term we show results 
for two parametrisations of the $K^{*}$ Regge trajectory:
the linear form (\ref{Kstar_trajectory_linear})
(see the lower solid lines) and
the ``square-root'' form (\ref{Kstar_trajectory_nonlinear}) (see the upper solid lines).
The absorption effects are included.}
\end{figure}

In Fig.~\ref{fig:4} we present the $K^{+}\pi^{-}K^{-}\pi^{+}$ 
invariant mass distributions for the continuum
$K^{*0}$-exchange contribution and 
the $f_{2}(1950)$-exchange contribution
for the parameters fixed by the WA102 data.
According to the same strategy 
as in the previous section
both contributions are considered separately, 
i.e. without possible interference effects between
the continuum $K^{*0} \bar{K}^{*0}$ 
and the signal $f_{2} \to K^{*0} \bar{K}^{*0}$ processes.
The calculations were done for $\sqrt{s} = 13$~TeV with
typical experimental cuts on $\eta_{M}$ (pseudorapidities) and
$p_{t,M}$ (transverse momenta) of centrally produced pions and kaons.
There are shown the results with an extra cut 
on momenta of leading protons
$0.17\;{\rm GeV} < |p_{y,p}| < 0.50\;{\rm GeV}$ that will be
applied when using the ALFA subdetector on both sides of the ATLAS detector.
We show results for larger (forward) pseudorapidities 
and without a measurement of outgoing protons
relevant for the LHCb experiment.
We can see that the distributions of both considered mechanisms
have maximum around $M_{K^{+}K^{-}\pi^{+}\pi^{-}} \simeq 2$~GeV;
therefore, the continuum term close to the threshold
may be misidentified as a broad $f_{2}(1950)$ resonance.
A clear difference is visible at higher values of the invariant mass 
of the $K^{+}K^{-}\pi^{+}\pi^{-}$ system.
The invariant mass distributions for the continuum contribution
is broader 
compared to the $f_{2}(1950)$ contribution
which we show for two cutoff parameters
$\Lambda_{f_{2},P} = 1.6$~GeV and 2.0~GeV.
For the continuum term we show results 
with the $K^{*}$ Regge trajectory both
for the linear form (\ref{Kstar_trajectory_linear})
(see the lower solid lines) and
the ``square-root'' form (\ref{Kstar_trajectory_nonlinear}) 
(see the upper solid lines).

In Table~\ref{tab:table2} we have collected integrated 
cross sections in nb for different experimental cuts
for the exclusive $K^{+}K^{-}\pi^{+}\pi^{-}$ production
via the intermediate $K^{*0} \bar{K}^{*0}$ states
including the contributions shown in Fig.~\ref{fig:diagrams}.
The ratio of the full and Born cross sections at $\sqrt{s}=13$~TeV
is approximately $\langle S^{2} \rangle \cong 0.19$ 
for the $K^{*0}$-exchange continuum contribution
and $\langle S^{2} \rangle \cong 0.18$ 
for the $f_{2}(1950)$-exchange contribution.
For the continuum, we used (\ref{Kstar_trajectory_linear}).
For the $f_{2}$ case we show the results for 
$\Lambda_{f_{2},P} = 1.6$~GeV (smaller cross sections)
and 2.0~GeV (larger cross sections).
\begin{table}[!ht]
\centering
\caption{The integrated cross sections in nb for the reaction 
$pp \to pp (K^{*0} \bar{K}^{*0} \to K^{+}\pi^{-}K^{-}\pi^{+})$
for the $f_{2}(1950)$ contribution (for $\Lambda_{f_{2},P} = 1.6 - 2.0$~GeV)
and the $K^{*0}$-exchange continuum contribution;
see Fig.~\ref{fig:diagrams}.
The results have been calculated for $\sqrt{s}=13$~TeV
and some typical experimental cuts on 
pseudorapidities and transverse momenta of produced pions and kaons.
The absorption effects were included here.}
\label{tab:table2}
\begin{tabular}{cl|c|c|c}\cline{3-4}&  & \multicolumn{2}{c|}{Cross sections (nb)} \\ 
\cline{1-4}
\multicolumn{1}{|c}{$\sqrt{s}$ (TeV)} & 
\multicolumn{1}{|c|}{Cuts} 
& $f_{2}(1950)$& Continuum   \\ 
\cline{1-4}
\multicolumn{1}{|c|}{13} &  
$|\eta_{M}| < 1.0$, $p_{t, M} > 0.1$~GeV
&18.6 -- 23.7 &  \;\,32.1  \\ 
\multicolumn{1}{|c|}{13} &  
$|\eta_{M}| < 2.5$, $p_{t, M} > 0.1$~GeV
&151.5 -- 190.0&  249.4  \\ 
\multicolumn{1}{|c|}{13} & 
 $|\eta_{M}| < 2.5$, $p_{t, M} > 0.2$~GeV
&56.5 -- 75.0&  109.2  \\ 
\multicolumn{1}{|c|}{13} &  
$|\eta_{M}| < 2.5$, $p_{t, M} > 0.2$~GeV, 
$0.17\; {\rm GeV} < |p_{y}| < 0.5\; {\rm GeV}$
&8.8 -- 11.7&  \;\,17.0  \\ 
\multicolumn{1}{|c|}{13} &  
$2 < \eta_{M} < 4.5$, $p_{t, M} > 0.1$~GeV
&58.6 -- 72.9&  \;\,93.1  \\  
\multicolumn{1}{|c|}{13} &  
$2 < \eta_{M} < 4.5$, $p_{t, M} > 0.2$~GeV
&23.3 -- 30.6&  \;\,43.1  \\ 
\cline{1-4}
\end{tabular}
\end{table}

Let us complete our analysis with the following remark.
We are assuming that the reaction $pp \to pp K^{*0} \bar{K}^{*0}$
is dominated by pomeron exchange,
for both the $f_{2}(1950)$ and continuum mechanisms,
already at $\sqrt{s} = 29.1$~GeV. 
Using this we have fixed some parameters of our model
and then have calculated the cross sections for the LHC.
But, the subleading reggeon-exchange contributions 
(e.g., $f_{2 \Reg} f_{2 \Reg}$-, $f_{2 \Reg} \Pom$-, 
$\Pom f_{2 \Reg}$-fusion processes) can also participate.
The inclusion of these subleading exchanges
would introduce many new coupling parameters
and form factors and would make a meaningful analysis of the WA102 data
practically impossible.
However, for the analysis of data from the COMPASS experiment,
which operates in the same energy range as previously
the WA102 experiment, it could be very worthwhile to study all
the above subleading exchanges in detail. 
Keep in mind that at high energies
and in the midrapidity region the subleading exchanges 
should give small contributions.
However, they may influence the absolute normalization 
of the cross section at low energies.
In general, our two mechanisms may have different production modes,
and therefore also different energy dependence of the cross section.
Therefore, our predictions for the LHC experiments
should be regarded rather as an upper limit.

\section{Conclusions}

In this paper, we have discussed diffractive production 
of $K^{*0} \bar{K}^{*0}$ system
in proton-proton collisions within the tensor-pomeron approach.
Two different mechanism have been considered,
central exclusive production of the $f_{2}(1950)$ resonance
and the continuum with the intermediate $K^{*0}$-meson exchange.
By comparing the theoretical results 
and the WA102 experimental data \cite{Barberis:1998tv}
we have fixed some coupling parameters 
and off-shell dependencies of an intermediate mesons.

We have shown that the continuum contribution alone,
taking into account the dominance of the $b$-type of the
$\Pom K^{*} K^{*}$ coupling,
describes the invariant mass spectrum obtained 
by the WA102 Collaboration reasonably well.
This is not the case for the $f_{2}(1950)$ meson
for which an agreement with the WA102 data
for the preferred type of couplings
$g^{(1)}_{\Pom \Pom f_{2}}$ and $g''_{f_{2} K^{*}\bar{K}^{*}}$
in the limited invariant mass range was found.
We have found that, in both cases, the model results 
are in better agreement with the WA102 data 
taking into account the tensor-vector-vector vertices (couplings) 
with the $\Gamma^{(2)}$ function rather than $\Gamma^{(0)}$ one.
This observation in our tensor-pomeron approach should be verified 
by future experimental results.
Hopefully this will be the case at the LHC.

In the calculation the absorptive corrections 
calculated at the amplitude level
and related to the proton-proton nonperturbative interactions 
have been included.
It is known that the absorption effects 
considerably change the shape of the distribution for $\phi_{pp}$, 
the azimuthal angle between the outgoing protons, 
and the shape of the distribution for ${\rm dP_{t}}$,
the difference of the transverse momentum vectors of the outgoing protons.
The $\phi_{pp}$ and ${\rm dP_{t}}$ dependences
for the $K^{*0} \bar{K}^{*0}$ system
was measured by the WA102 Collaboration.
We have reproduced these results fairly well in our model
including absorption effects.
The final distributions in these variables for both mechanisms considered
are very similar to each other.

Assuming that the WA102 data are already dominated by pomeron exchange,
we have calculated the cross sections 
for the reaction $pp \to pp K^{*0} \bar{K}^{*0}$
for experiments at the LHC
imposing cuts on pseudorapidities 
and transverse momenta of the pions and kaons from the decays
$K^{*0} \bar{K}^{*0} \to (K^{+}\pi^{-})(K^{-}\pi^{+})$.
The distributions of the invariant mass 
of the $K^{+}\pi^{-}K^{-}\pi^{+}$ system have been presented.
We should keep in mind that both considered mechanisms 
have a maximum around $M_{K^{+}K^{-}\pi^{+}\pi^{-}} \simeq 2$~GeV, 
thus a broad enhancement (at least part of it) 
in this mass region 
can be misidentified as the $f_{2}(1950)$ resonance.
Furthermore, a similar behaviour of the 
continuum and $f_{2}$ production processes 
makes an identification of a hypothetical tensor-glueball state 
in this reaction rather difficult.

We have found in this paper that the diffractive processes
leads to a cross section 
for the $K^{*0} \bar{K}^{*0} \to K^{+}\pi^{-}K^{-}\pi^{+}$ production
more than one order
of magnitude larger than the corresponding cross section
for the $\phi(1020)\phi(1020) \to K^{+}K^{-}K^{+}K^{-}$ processes
considered in \cite{Lebiedowicz:2019jru}.
Our predictions can be tested by all collaborations
(ALICE, ATLAS, CMS, LHCb) working at the LHC.
A measurable cross section for the exclusive process
$pp \to pp (K^{*0} \bar{K}^{*0} \to K^{+}\pi^{-}K^{-}\pi^{+})$
should provide an interesting challenge
to check and explore.

\acknowledgments
\vspace{-0.5cm}
I am indebted to Antoni Szczurek and Otto Nachtmann for useful discussions.
This work was partially supported by
the Polish National Science Centre under Grant
No. 2018/31/B/ST2/03537.

\bibliography{refs}

\end{document}